# Animer une base de connaissance : des ontologies aux modèles d'I.A. générative.

Des systèmes experts et du web sémantique à l'I.A. générative : approches « model-driven » et « data-driven » en études aréales.

## Peter Stockinger

Cnrs - Centre Internet et Société (C.I.S.)

Inalco - Pluralité des Langues et des Identités (PLIDAM)

Lambach – Paris 2025

# TABLE DES MATIERES

## RESUME

Dans un contexte où les sciences sociales et humaines expérimentent des cadres analytiques non anthropocentriques, cet article propose une lecture sémiotique (structurale) de l'hybridation entre l'IA *symbolique* et l'IA *neuronale* (ou *sous-symbolique*) à partir d'un champ d'application : le design et l'usage d'une base de connaissance pour les études aréales. Nous décrivons l'écosystème *LaCAS – Archives ouvertes en études linguistiques et culturelles* (thésaurus ; ontologie RDF/OWL ; services LOD ; moissonnage ; expertise ; publication), déployé à l'Inalco (Institut national des langues et civilisations orientales) avec l'environnement logiciel Okapi (Open Knowledge and Annotation Interface) de l'Ina (Institut national de l'audiovisuel), qui compte aujourd'hui environ 160 000 ressources documentaires et une dizaine de macro-domaines regroupant plusieurs milliers d'objets de connaissance. Nous illustrons cette approche à partir du domaine de connaissance « Langues du monde » (~540 langues) et de l'objet de connaissance « Quechua (langue) ».

Sur cette base, nous discutons de l'intégration contrôlée d'outils neuronaux, plus précisément d'outils génératifs, dans le cycle de vie d'une base de connaissance : aide à la localisation/qualification des données, extraction et agrégation d'index, suggestion et test de propriétés, génération dynamique de fichiers, ingénierie de prompts contextualisés alignés sur l'ontologie. Nous esquissons un écosystème d'agents spécialisés capables d'animer la base de données tout en respectant ses contraintes symboliques, en articulant des méthodes *model-driven* et *data-driven*.

L'angle non-humain n'hypostase pas une machine « pensante » ; il remet en question nos catégories d'analyse : externalisation de l'agentivité dans les artefacts, circulation entre des systèmes de signes hétérogènes, redistribution des frontières du « faire-sens » entre humains et non-humains (techniques et vivants). Dans le sillage des changements épistémologiques contemporains (sémiotique, Umwelt, affordances, cognition incarnée et située), nous concevons la base de connaissance comme un milieu sémiotique où sont expérimentés des procédés de modélisation, de représentation, de traduction, d'alignement et de pilotage méthodologique.

L'argumentation de cet article suit un fil conducteur du symbolique à l'hybride : après l'introduction (le cadre « Penser le non-humain » et le positionnement sémiotique), nous revenons à l'IA symbolique et aux systèmes experts (rappels et rôle central de la base de connaissance), puis au web sémantique (RDF/OWL ; LOD) pour décrire comment les graphes de connaissance soutiennent le raisonnement et la publication. Nous présentons ensuite la plateforme LaCAS et sa base de connaissance (thésaurus, macro-domaines de connaissance, objets de connaissance), avant d'aborder un cas spécifique (« Langues du monde » / Quechua) qui nous fait prendre conscience de la richesse du modèle conceptuel et de la documentation structurée qui représente l'expertise d'un objet de connaissance. Enfin, nous nous ouvrons à l'hybridation symbolique ↔ neuronale (techniques modèle-centrées / données-centrées, prompts et écosystèmes d'agents), qui justifie la transition « des ontologies aux modèles d'IA générative ».

Concrètement, à partir du cas LaCAS, nous montrons comment un modèle conceptuel (classes/propriétés) et un modèle de données textuelles peuvent structurer des *observatoires vivants* d'objets de connaissance, des cartographies comparatives et, via des prompts bien contraints, accélérer la mise à jour raisonnée du graphe de connaissance sans perdre en rigueur critique. Nous concluons en présentant les limites (biais de source, explicabilité, dépendance vis-à-vis des référentiels) et les perspectives d'une telle hybridation pour « penser le non-humain » dans des écosystèmes de connaissances véritablement partagés.

## ABSTRACT

**Animating a Knowledge Base: From Ontologies to Generative AI Models**
*From Expert Systems and the Semantic Web to Generative AI: Model-Driven and Data-Driven Approaches in Area Studies*

In a context where the social sciences and humanities are experimenting with non-anthropocentric analytical frames, this article proposes a semiotic (structural) reading of the hybridization between *symbolic* AI and *neural* (or *sub-symbolic*) AI based on a field of application: the design and use of a knowledge base for area studies. We describe the *LaCAS* ecosystem – *Open Archives in Linguistic and Cultural Studies* (thesaurus; RDF/OWL ontology; LOD services; harvesting; expertise; publication), deployed at Inalco (National Institute for Oriental Languages and Civilizations) in Paris with the Okapi (Open Knowledge and Annotation Interface) software environment from Ina (National Audiovisual Institute), which now has around 160,000 documentary resources and ten knowledge macro-domains grouping together several thousand knowledge objects. We illustrate this approach using the knowledge domain "Languages of the world" (~540 languages) and the knowledge object "Quechua (language)".

On this basis, we discuss the controlled integration of neural tools, more specifically generative tools, into the life cycle of a knowledge base: assistance with data localization/qualification, index extraction and aggregation, property suggestion and testing, dynamic file generation, and engineering of contextualized prompts (generic, contextual, explanatory, adjustment, procedural) aligned with a domain ontology. We outline an ecosystem of specialized agents capable of animating the database while respecting its symbolic constraints, by articulating *model-driven* and *data-driven* methods.

The non-human angle does not hypostasize a "thinking" machine; it challenges our categories of analysis: externalization of agency in artifacts, circulation between heterogeneous sign systems, redistribution of the boundaries of "meaning-making" between humans and non-humans (technical and living). In the wake of contemporary epistemological changes (semiotics, Umwelt, affordances, embodied and situated cognition), we conceive of the knowledge base as a semiotic milieu where procedures of modeling, representation, translation, alignment, and methodological steering are experimented with.

The argument of this article follows a thread from the symbolic to the hybrid: after the introduction (the "Thinking the non-human" frame and the semiotic positioning), we return to symbolic AI and expert systems (reminders and the central role of the knowledge base), then to the semantic web (RDF/OWL; LOD) to explain how knowledge graphs support reasoning and publication. We then present the LaCAS platform and its knowledge base (thesaurus, knowledge macro-domains, knowledge objects, knowledge hubs), before addressing a specific case ("Languages of the world" / Quechua) that makes us aware of the richness of the conceptual model and structured documentation that represents the expertise of a knowledge object. Finally, we open up to symbolic ↔ neural hybridization (model-centric/data-centric practices, prompts, and agent ecosystems), which justifies the transition "from ontologies to generative AI models."

Concretely, using the LaCAS case, we show how a conceptual model (classes/properties) and a textual data model can structure knowledge hubs and living observatories of knowledge objects, comparative cartographies and, via well-constrained prompts, accelerate the reasoned updating of the knowledge graph without losing critical rigor. We conclude by presenting the limitations (source bias, explainability, dependence on repositories) and prospects of such hybridization for "thinking the non-human" in truly shared knowledge ecosystems.

# 1) INTRODUCTION

Pour contribuer à la problématique générale de ce numéro, je m'intéresserai à un domaine de recherche appliquée qui est celui de la conception et de la réalisation de *bases de connaissance*. Les bases de connaissance constituent une des applications majeures en intelligence artificielle. Ce sont des systèmes informatiques dont l'output attendu (sous forme, par exemple, d'une réponse à une question, d'une aide à une prise de décision ...) se fonde sur un input qui est formé de données qualifiées et interprétées par de modèles conceptuels représentant les connaissances d'un acteur ou, plutôt, d'une communauté d'acteurs.

Le terme « intelligence artificielle » est aujourd'hui employé pour désigner des problématiques, des approches et des recherches technologiques qui sont très variées. Schématiquement, on distingue deux approches complémentaires :

1. L'*approche symbolique* (« intelligence artificielle symbolique » ou *I.A. symbolique*). Cette approche a recours à des techniques de représentation et de modélisation explicites de connaissances d'un domaine donné et à la logique formelle (de prédicats, notamment) pour résoudre un ensemble de problèmes tels que ceux de la reconnaissance d'un objet, de la classification d'un objet, de la planification d'une action, de la simulation d'une situation. Historiquement, les recherches consacrées aux bases de connaissance font partie de cette approche symbolique.
2. L'*approche neuronale* ou *connexionniste*[1]. Cette approche recourt aux techniques de représentation numérique (comprise ici au sens d'une représentation *vectorielle*) de données et de méthodes d'entraînement de réseaux neuronaux artificiels afin de résoudre le même genre de problèmes que l'IA symbolique, mais sans passer par le carcan d'un modèle – d'une *ontologie* – explicite du ou des domaines concernés. Cette approche se différencie dans des domaines plus particuliers : génération de données textuelles lato sensu (= domaine de l'IA générative) ; prédiction ou simulation de situations futures à partir de données existantes (= domaine de l'IA prédictive) ; capacité d'agir en fonction d'un objectif, d'un plan et des contraintes données (= domaine de l'IA agentive).

Très généralement parlant, l'intelligence artificielle englobent la diversité des technologies qui, comme souligné à maintes reprises par de nombreux experts et spécialistes (cf., à titre d'exemple, Mira 2008 ; Barraud 2020), concernent directement les capacités cognitives typiquement humaines telles que la perception, l'évaluation, la hiérarchisation et la reconnaissance de régularités naturelles ou sociales caractérisant un milieu ; la thématisation de domaines d'expérience ou d'expertise ; la conception et la spécification de programmes d'actions en fonction d'un milieu et d'un objectif donnés ; la contextualisation, la qualification et la classification de données documentant un domaine d'expérience ou d'expertise.

---

[1] On parle également de l' « IA sous-symbolique » (cf., par exemple, Ilkou, E., & Koutraki, M. 2020)

Analogues à la bionique, ces technologies s'inspirent des capacités (ou compétences) que les êtres humains (et, au-delà, les êtres vivants à différents niveaux) ont développées pour répondre à leurs besoins et à leurs désirs dans leur milieu. Il ne s'agit pas obligatoirement de produire un « homme artificiel » (ou un « *être* artificiel ») mais plutôt de simuler des recherches de solution de tâches problématiques (ou, plutôt, ensembles de tâches) dans un milieu donné en mobilisant des artefacts qui s'appuient sur lesdites compétences ou, pour être plus précis, sur des *modèles* ou *représentations* de ces compétences. Ainsi, un robot qui est utilisé dans l'industrie pour assembler certaines pièces selon une succession automatisée d'instructions d'action représente une solution technologique à un travail manuel humain dans un milieu de production industrielle particulière. Le moteur de recherche qui propose à un utilisateur particulier une liste de réponses pouvant satisfaire les désirs de ce dernier, représente une solution technologique du travail d'un documentaliste ou d'une personne spécialisée en la collecte et la restitution d'informations pertinentes.

Il nous paraît important de rappeler ces évidences pour trois raisons. D'abord, les solutions technologiques s'inspirent, bien sûr, de données existantes (ici, de solutions cognitives historiquement et culturellement *situées*) mais ne prétendent pas à priori de devenir des sortes de machines qui sentent ou qui pensent « comme un être humain » ou (« comme un être vivant tout court »). Cette prétention peut faire partie d'un programme de recherche, mais ne le fait pas obligatoirement. Lorsqu'on parle d'intelligence artificielle, il faut donc considérer avec précaution l'usage du terme "intelligence" et de sa ou, plutôt, ses significations implicites.

Une deuxième raison est que l'explication et la simulation des dispositions ou des compétences « typiquement » humaines peuvent s'inscrire dans une logique dictée par deux objectifs certes liés, mais néanmoins distincts. Il peut s'agir d'une recherche fondamentale sur les « mécanismes » (*physiques,* mais pas nécessairement matériels !) qui sous-tendent ces dispositions et ces compétences, ou sur l'émergence et l'évolution des régularités qui les caractérisent. Ils peuvent également se concentrer sur des solutions à des défis pratiques et sociétaux au sens large du terme, tels que libérer les humains de la corvée du travail manuel monotone, augmenter l'efficacité du travail normalement effectué par des agents humains ou, d'une manière très générale, contribuer davantage à la satisfaction d'un désir ou d'un besoin anthropique.

Une troisième raison est que toute recherche en intelligence artificielle est évidemment tributaire des connaissances existantes sur la ou les problématiques qui constituent son champ de recherche. Elle est ainsi tributaire des approches et des visions (ou « paradigmes scientifiques ») qui tentent d'expliquer structure et fonctionnement des dispositions ou des compétences « typiquement » humaines. Autrement dit, les modèles utilisés en intelligence artificielle pour simuler les capacités cognitives mentionnées ci-dessus sont plutôt l'expression d'un certain *point de vue* (plus ou moins explicite, plus ou moins partagé…) et d'un certain *état de savoir* (toujours en principe révisable) sur ces capacités.

Ces précisions montrent qu'il est tout à fait compréhensible que l'intelligence artificielle soit un puissant stimulateur de spéculations imaginatives sur l'évolution (post-humaine voire non humaine) des humains et de leur milieu. Cependant, les applications de cette recherche (sous la forme, par exemple, de toutes sortes d'artefacts qui enrichissent et augmentent les capacités humaines actuelles) peuvent également servir d'indices révélateurs des représentations culturelles et historiques qui servent de standards (i.e. de références) à une civilisation (à notre civilisation) pour réfléchir sur l'homme et la condition humaine. Toute révolution culturelle (toute innovation technique ou technologique) a comme conséquence des changements plus ou moins significatifs qui affectent la vision que possède un acteur ou une population d'acteurs (formant une « civilisation ») de ce que c'est l'humain (et, par conséquence, de ce que c'est le « non-humain ») et du milieu qui constitue son monde de vie (et, par conséquence, du milieu qui caractérise, pour parler ainsi, l'anti-monde de vie, voire un monde de vie non-humaine). Nous interprétons ainsi les débats autour de l'intelligence artificielle en particulier et des technologies qui s'appuient sur les dispositions et capacités cognitives humaines soit pour résoudre des problèmes pratiques, soit pour « enrichir » ou « augmenter » les dispositions et capacités actuelles. Enfin, il est difficile de voir comment la recherche « faite par l'homme » (qu'elle implique ou non l'intelligence artificielle) pourrait dépasser les limites du langage, pour citer Ludwig Wittgenstein.

Nous nous intéresserons, dans cet article, aux deux approches dans les recherches en intelligence artificielle citées au début de notre introduction, à savoir, à l'approche symbolique et à l'approche neuronale.

Cet article est basé en grande partie sur nos expériences et recherches qui, depuis déjà fort longtemps, ont régulièrement rencontré des problématiques, méthodes et approches relevant de l'intelligence artificielle. En effet, c'est depuis le milieu des années 80 ((Stockinger 1985, 1986, 1987, 1994) ; voir aussi (Fargues 1991), lorsque nous travaillions au Cnrs, que nous étions intéressés aux questions relatives à l'extraction, à la modélisation et à la représentation des connaissances. Ces précisions un peu biographiques pour deux raisons :

– d'abord, nous nous appuierons sur un projet récent d'une *plateforme sémantique* mise en place à l'Inalco (Institut National des Langues et Civilisations Orientales) qui intègre une approche conceptuelle et informatique très particulière pour traiter des données de recherche en *études aréales* ;
– ensuite, nous partirons d'une compréhension dite symbolique de l'intelligence artificielle tout en nous intéressant aux apports de l'intelligence artificielle générative, prédictive et agentive.

L'observation qui nous a toujours intrigué est la suivante : les problématiques que traitent aussi bien l'intelligence artificielle symbolique que l'intelligence artificielle neuronale, nous sensibilisent à l'importance d'une *expertise*, d'un *savoir-faire* que nous qualifierons de *sémiotique* au sens large du terme sans penser obligatoirement à une « école » ou à un « courant » particulier. Citons les deux questions que nous discuterons plus en avant dans notre contribution :

1. La question de la qualification, de la constitution et du traitement de *corpus de données multimodales*, voire hétérogènes qui, ensemble, documentent un domaine donné.
2. La question de l'élaboration de *modèles* – d'ontologies – représentant d'une manière explicite le savoir et savoir-faire que possède un acteur (ou une population d'acteurs) sur un domaine donné.

Je reviendrai sur ces deux problématiques tout au long de cet article. Le chapitre 2 est consacré à un bref rappel historique des systèmes experts qui constituent un champ historique d'application majeur en intelligence artificielle symbolique. Le chapitre 3 présente rapidement l'évolution de l'intelligence artificielle symbolique dans sa version du web sémantique, i.e. de la tentative de transformer l'ensemble du web en une base de connaissance. Les chapitres 4 et 5 présentent plus en détail l'écosystème d'une base de connaissance en prenant comme exemple le projet LaCAS de l'Inalco. Le chapitre 6 est réservé à une discussion plus approfondie d'un domaine de connaissance intitulé *Langues du monde*. Enfin, le chapitre 7 s'intéresse au « mariage » des deux approches et aux conséquences d'un tel rapprochement pour l'évolution des systèmes à base de connaissance.

## 2) L'APPROCHE SYMBOLIQUE EN INTELLIGENCE ARTIFICIELLE

Historiquement parlant, l'intelligence artificielle désigne des recherches en même temps théoriques et appliquées qui se réfèrent à la logique mathématique (par exemple, à la logique des prédicats, aux logiques modales ou aux diverses logiques d'action) pour formaliser le raisonnement humain dans certains domaines d'expertise bien circonscrits.

Les recherches en IA symbolique ont eu des répercussions importantes sur des approches et disciplines émergentes telles que le *TAL* (traitement automatique des langues) et la *cognitique*. Le TAL comprend un ensemble d'approches et de méthodes d'analyse et d'annotation de corpus de données multilingues, de compréhension sémantique de textes, de génération de textes, de traduction d'énoncés et de textes entiers ou encore de fouille statistique et thématique (« topique ») dans des larges bases de données textuelles. La cognitique – aussi appelée « analyse conceptuelle » ou « information design » (Stockinger 2018) - s'intéresse aux problèmes relatifs à l'élaboration des bases de connaissance, d'extraction et de modélisation de connaissances à partir de données textuelles lato sensu (données écrites ou orales, mais également données visuelles et audiovisuelles, données polysensorielles, etc.)[2].

L'un des domaines les plus emblématiques de recherche et de développement en IA symbolique est celui des *systèmes experts*. Un système expert est (ou est censé être) capable de simuler le raisonnement humain dans des domaines d'expertise spécifiques. Les exemples historiques les plus connus sont les systèmes experts dans les domaines industriel, médical et juridique. Cependant, on trouve des exemples de systèmes experts dans pratiquement tous les domaines de connaissance. L'utilisation concrète de systèmes experts est envisagée tantôt au sens d'un outil d'aide à l'apprentissage et à la décision, tantôt d'une manière plus présomptueuse au sens d'un « robot symbolique » qui est censé remplacer l'expert humain pour résoudre une gamme de problèmes qui font partie d'un domaine d'expertise (cf., par exemple, Ermine 2000).

En fait, parmi les premiers systèmes experts en France figuraient ceux conçus et réalisés en archéologie par Jean-Claude Gardin et ses collaborateurs (Gardin 1987, 1989, 1990)[3]. Un des enjeux de ces recherches était d'aider les spécialistes à prendre des décisions concernant l'identification, la datation et la classification d'objets excavés sur des sites archéologiques, voire de sites archéologiques entiers.

Un système expert comprend deux grands modules : le module appelé *base de connaissance* et le module appelé *moteur d'inférence* qui exploite la base de connaissance pour résoudre un problème qui lui est posé. Une base de connaissance est constituée d'un ensemble de *faits*. Un fait est une *proposition* (ou *énoncé générique*) qui prédit l'existence d'une caractéristique (d'une régularité) significative dans le domaine d'expertise qui constitue l'objet d'un système expert, ainsi que toute relation significative entre une caractéristique et d'autres caractéristiques. En médecine, par exemple, pour diagnostiquer une maladie, on s'appuie sur une sémiologie de symptômes : température corporelle,

---

[2] Cf. l'ouvrage de Bench-Capon, T. J., *Knowledge representation.* An approach to artificial intelligence, Amsterdam, Elsevier, 2014

présence de douleurs, état de la peau, etc. Toute maladie, selon un standard médical donné, peut ainsi être caractérisée par un ensemble de symptômes. Ces symptômes sont documentés par des données textuelles au sens large du terme : descriptions verbales d'experts, représentations photographiques, schémas, données épidémiologiques, données de santé publique.

Cela dit, réduire un *fait* à une affirmation (ou à une négation) est une simplification excessive qui ne tient pas compte de la complexité *indexicale*, *topique*, *agentive* ainsi qu'*empirique* qui entoure cette opération cognitive[4] :

- *complexité indexicale* signifie que toute expertise est conditionnée par la *position* d'un expert dans un champ d'experts (d'expertises) disponibles ;
- *complexité topique* signifie que toute expertise thématise son objet dans un *horizon de sens* qui est conditionné par le point de vue sémantique adopté servant de standard pour « parler » de l'objet en question ;
- *complexité agentive* renvoie aux *dispositions subjectives* de l'expert (i.e., à son *agentivité*, autrement dit, à sa compétence cognitive et potestative, à sa volonté, à son état émotionnel ou encore à sa confiance) de se servir d'un standard pour développer son expertise ;
- *complexité empirique* renvoie à l'*exhaustivité*, à la *représentativité* ou encore à l'*actualité* d'une expertise.

Afin de mieux comprendre le discours expert, c'est-à-dire la diversité existante des experts et des discours experts dans un domaine spécifique, de nombreuses recherches en intelligence artificielle symbolique ont commencé à aborder des questions théoriques très similaires à celles qui ont dominé la sémiotique structurale dans les années 1970 et 1980 - même si ces deux « disciplines » n'ont jamais réellement entretenu des rapports directs et systématiques. Mentionnons, à titre d'exemple seulement, les recherches sur la *multi-expertise* comprise comme le « produit » (toujours révisable) d'une communauté d'experts, sur les *rôles actantiels* organisant la communauté d'experts engagés dans un projet d'expertise ou encore sur les *genres sémiotiques* (topiques, discursifs, textuels à proprement parler et médiatiques) servant de standards ou de « normes » de production ainsi que d'appropriation et d'utilisation de données textuelles lato sensu. En effet, l'*IA symbolique* aurait pu – et aurait dû – être un domaine de recherche particulièrement adapté à la vision du *langage* développée par A.J. Greimas et d'autres structuralistes, mais, comme l'histoire nous l'a montré, cette voie n'a été suivie que très timidement par la communauté scientifique concernée[5].

Considérons rapidement la deuxième partie d'un système expert, à savoir le module de solution de problème, appelé *moteur d'inférence*. Pour prendre notre exemple d'un système expert médical, le moteur d'inférence est censé simuler le *raisonnement* d'un *expert* qui comprend, sur la base de son savoir et de son expérience, l'état d'une personne représentant un ensemble de symptômes et comment l'aider. Dans ce cas précis, ce module doit, entre autres :

- fournir une aide à la décision si une personne présentant certains symptômes a effectivement contracté une maladie donnée ;
- proposer une probabilité raisonnée si une personne présentant certains symptômes a contracté une maladie donnée ;

---

[4] cf. également l'évaluation critique de l'IA symbolique par Santoro, A. et al (2021)
[5] cf., cependant, Stockinger et al (1985, 1986) ; Arnold, M. (1989) ; Meunier, J.-G. (1992) ; Stockinger (1994) ; Anderson 1997 ; Barron, T. et al (1999) ; Valdez, J. L. C. et al (2024)

–   proposer une évaluation raisonnée de la probabilité qu'une personne soit porteuse d'une ou plusieurs autres maladies.

Bien entendu, un système expert qui intègre une base de connaissance empiriquement riche, mise à jour régulièrement par des nouvelles données, prenant en compte le contexte (historique, social, psychologique, médical à proprement parler) pourra fournir bien d'autres renseignements et conseils. Il pourra, par exemple, offrir des conseils thérapeutiques ou encore procéder à un tri argumenté entre malades représentant un profil épidémiologique typique et ceux dont le profil pose davantage de problèmes.

Le processus du raisonnement se différencie en des *opérations cognitives* d'inférence, de comparaison, de classification, d'élimination d'alternatives, de conditionnalité, etc. qu'on formalise sous forme de *règles logiques* (d'inférence). Un *algorithme*, à son tour, intègre ces règles en un tout cohérent pour fournir, en un nombre limité d'étapes, une solution à un problème (ou, plutôt, à une classe de problèmes). La partie « moteur » ou « raisonneur » comprend ainsi à la fois des *règles générales* (par exemple d'inférence, de classification, de « chaînage en avant/en arrière d'instructions », …) qui servent à l'exploitation de bases de connaissance les plus diverses, et des *règles* dont la portée est *contextuelle*, voire simplement *« ad hoc »*. Les règles appartenant à cette deuxième catégorie sont conditionnées par la spécificité du domaine d'expertise que représente une base de connaissance ainsi que les objectifs qu'un système expert est censé satisfaire.

Conception et réalisation d'un système expert – de sa base de connaissance peuplée de données interprétées et de son moteur d'inférence – sont des tâches longues et complexes comprenant, entre autres :

1.   la constitution d'un corpus de données textuelles qui documentent un savoir et savoir-faire qu'on cherche à expliciter ;
2.   l'élaboration d'un modèle conceptuel du domaine d'expérience ou d'expertise à l'aide de modèles conceptuels et formels ;
3.   l'identification, l'extraction, la classification et l'annotation des informations pertinentes contenues dans le corpus de données ;
4.   la formulation de règles contextuelles et/ou ad hoc et l'adaptation du moteur d'inférence ;
5.   la réalisation d'un ou, plutôt, de plusieurs prototypes avant celle du système en « grandeur réelle » ;
6.   l'intégration du système expert dans un environnement de travail et la formation des utilisateurs.

Ce sont des projets de recherche et de développement fort complexes qui mobilisent des compétences variées. Quelques enjeux récurrents qu'on rencontre régulièrement ici concernent[6] :

1.   la qualité du modèle conceptuel lui-même et sa *reconnaissance* par la communauté d'experts (enjeu qui renvoie à la problématique de la *multi-expertise* déjà citée) ;
2.   la qualité (fiabilité) et la couverture empirique des données et l'enrichissement continu d'une base de connaissance ;
3.   le processus d'extraction à proprement parler qui, à partir d'un certain seuil, ne peut plus se faire d'une manière *manuelle* (i.e. sous forme d'une annotation structurée des données d'une base de connaissance) et qui doit donc intégrer des méthodes

---

[6] Cf. à ce sujet la thèse très intéressante de Matthieu Bellucci, intitulée *Symbolic approaches for explainable artificial intelligence* (Bellucci 2023)

automatique (par exemple, de fouille de textes et de données) qui posent à leur tour un évident problème de fiabilité ;

4. le processus du suivi (de la mise à jour…) d'une part, du corpus des données et, d'autre part, de la base de connaissance.

Voilà succinctement les systèmes experts qui constituent une des applications principales de ce qu'on entendait « à l'époque » - dans les années 80 et 90 — par *intelligence artificielle*.

## 3) LE WEB SEMANTIQUE

L'IA symbolique est aujourd'hui représentée par les recherches et les applications dans le domaine de la *sémantique du web*, d'une part, et le *« linked open data »* (i.e. le LOD ; en français, les *données ouvertes liées*), d'autre part. On identifie l'ensemble des recherches, méthodes, outils et applications concernés par l'expression *Web sémantique* (ou *toile sémantique*) – expression créée déjà en 1999 par Tim Berners-Lee, un informaticien et physicien anglais qui en est une des figures les plus emblématiques[7].

À l'instar des systèmes experts, le principe directeur consiste ici à transformer le web en une *vaste base de connaissance* fondée sur des modèles sémantiques explicites pour décrire schématiquement une grande diversité de domaines et d'objets de connaissance afin de catégoriser, de classer les données en ligne. Le projet le plus abouti et le plus connu à ce jour est *Wikidata*. Wikidata est une immense base de connaissance qui transforme les entrées de l'encyclopédie collective Wikipédia en données structurées, interprétées et interconnectées, tout en bénéficiant des contributions d'une multitude de « producteurs » de connaissances.

Ainsi, par exemple, une langue naturelle telle que le *français* compris comme une *entité* (ou « *élément* ») *Wikidata* est un *objet de connaissance* caractérisé par un ensemble de *propriétés* qui spécifient son appartenance à une famille linguistique, ses particularités grammaticales et sociolinguistiques, son histoire, son rayonnement géographique, son enseignement, sa valorisation. Les propriétés qui définissent une entité Wikidata sont des *énoncés* ou des *propositions* qui affirment que l'entité possède une caractéristique particulière. Parallèlement, les informations sur l'entité *français* sont « enrichies » par les contributions d'une grande variété de « producteurs externes (i.e. externes au projet Wikimédia) » : *encyclopédies* (telles que l'Encyclopædia Universalis, l'Enciclopedia Italiana, l'Encyclopædia Britannica, etc.) ; *bibliothèques* (telles que la Bibliothèque nationale de France, la Library of Congress, la Deutsche Nationalbibliothek, etc.) ; *bases de données* et *thésaurus* (Glottolog, Linguasphere, WALS, OpenAlex, Internet Archive, …), etc. Ces contributions relèvent de la partie LOD (*Linked Open Data*) et, transformées en un ensemble de propriétés, déclarent que pour une entité Wikidata donnée, des informations sont disponibles dans un certain nombre de sources d'information (de « producteurs » de connaissances externes). Une entité Wikidata telle que le *français* est ainsi présentée comme un *hub de connaissances et d'informations* structuré, ouvert et dynamique qui peut être interrogé, exploré et exploité en fonction des besoins et des objectifs d'un utilisateur. Un hub de connaissances et d'informations est un *graphe (de connaissances) instancié* qui possède sa propre configuration. Il peut être intégré tel quel ou moyennant certaines modifications dans des bases de connaissance autres que Wikidata auxquelles il apporte aussi bien sa *structure* (le modèle des propriétés qui le définissent) que les *informations* collectées et agrégées pour comprendre, analyser et utiliser une entité donnée. Nous discuterons un exemple concret dans le chapitre 6.

Techniquement parlant, le Web sémantique repose sur un *format standard* pour décrire et traiter les données, ainsi que sur des *protocoles* pour y accéder. Le format de base s'appelle RDF (*Resource Description Framework*). Ce format est complété par des formats plus

---

[7] Cf. Renaud Fabre (2017) sur les enjeux technologiques et sociaux plus généraux de la « connaissance »

spécialisés parmi lesquels le plus important est l'OWL (*Web Ontology Language*), qui est un langage utilisé pour développer des *ontologies d'un domaine de connaissance,* i.e., des modèles qui expliquent, à un certain niveau de généricité, l'organisation et le fonctionnement d'un domaine du genre *Langues du Monde*, pour rester avec notre exemple du *français* comprise comme *entité Wikidata*.

Sans entrer dans des détails techniques et formels trop poussés, le RDF et, plus encore, l'OWL sont des modèles de *graphes*, i.e., des modèles formés par des paires de *sommets* (ou « nœuds ») et d'*arêtes* (ou « chemins »), une arête étant composée d'une *paire de sommets* (ou nœuds). La description d'un domaine de connaissance au format RDF s'entend comme une agrégation d'une ou d'un ensemble d'*énoncés* ou de *propositions* concernant un objet que nous souhaitons décrire. La structure *de base* de l'énoncé est très simple et se compose de deux nœuds et d'une arête qui transforme les deux nœuds en un *nœud source* (*d'origine*) et un *nœud cible*. Le *nœud source* est appelé « sujet », le *nœud cible* est appelé « objet » et l'arête orientée est appelée « prédicat ». Une déclaration RDF est donc composée d'un triplet et d'une description « conforme à RDF » (cf. figure 1). Autrement dit, une description qui fait référence au cadre du web sémantique, est un ensemble de triplets qui forment un graphe.

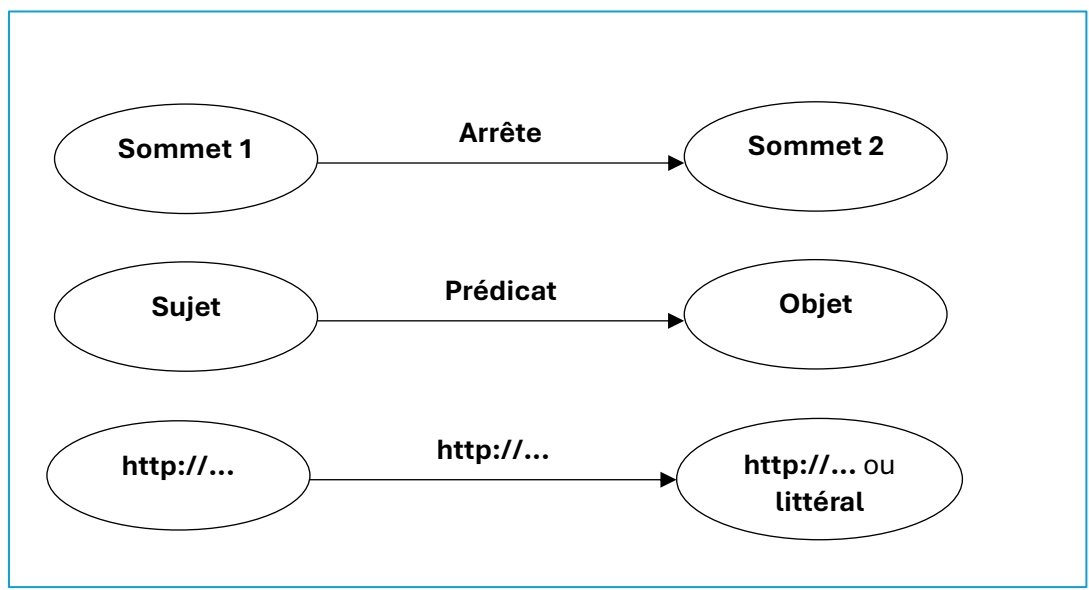

(Figure 1)

Cette approche consistant à produire des descriptions schématiques de domaines de connaissance est apparue dans les années 1950 et 1960 et s'est popularisée à la fin des années 1960 grâce à une série d'articles de recherche d'Alan M. Colins et M. Ross Quillian (1969) en psychologie cognitive consacrés à la mémoire sémantique. Dans les années 1970 et 1980, des propositions similaires ont été avancées par des chercheurs en intelligence artificielle, en traitement automatique du langage et en sciences cognitives. Citons, à titre d'exemple, la *théorie des cadres* (*frame theory*, en anglais) de M. Minsky (1974, 1986) et la *théorie des scripts* (ou scénarios) développée par Roger Schank et Robert P. Abelson (1975). Aujourd'hui, les réseaux sémantiques ou conceptuels sont intégrés et largement utilisés dans les langages ontologiques du web sémantique. L'une des contributions les plus importantes dans ce domaine est celle de John Sowa (1984), auteur d'une théorie des *graphes conceptuels* appliqué au TAL. Cette théorie a joué et continue de jouer un rôle central dans la recherche en intelligence artificielle et en cognitive, en particulier dans le domaine de la représentation des connaissances (voir également Stockinger 1994).

Une distinction fonctionnelle centrale qu'on rencontre dans les recherches sur la sémantique du web est celle entre les *données* et les *métadonnées*. Les données comprennent les *objets* d'une description. Les métadonnées comprennent les éléments qui *servent à décrire* un objet. L'objet de description auquel on pense ici d'abord est la *donnée textuelle (lato sensu)* comprise comme *document* qui renseigne, qui fournit des informations sur un domaine particulier. Cet exemple est illustré par des activités qui possèdent une longue tradition et représentent le travail d'expertise d'une grande diversité de métiers de recherche. On peut distinguer entre plusieurs grands types de métadonnées (documentaires) qui regroupent les propriétés utilisées pour produire une expertise. Il s'agit notamment :

- des *métadonnées techniques*, qui sont utilisés pour identifier et décrire le format médiatique d'une donnée textuelle (une image, un texte, des données multisensorielles, etc.) ;
- des *métadonnées paratextuelles*, qui sont utilisées pour décrire le cadre auctorial d'une donnée textuelle ;
- des *métadonnées descriptives*, qui sont utilisées pour identifier et expliquer le contenu pertinent d'une donnée textuelle ;
- des métadonnées *évaluatives* ou *appréciatives*, qui servent à produire un métadiscours critique sur la donnée textuelle elle-même, sur le contenu qu'elle fournit, sur ses origines, son cycle de vie, etc. ;
- des *métadonnées bibliographiques et webographiques*, qui servent à insérer une donnée textuelle dans un champ plus large de production textuelle ;
- des *métadonnées génétiques*, qui considèrent l'histoire et le cycle de vie d'une donnée textuelle.

La capture d'écran ci-dessous (figure 2) montre une interface de description d'une ressource audiovisuelle. Elle fait partie de la plateforme sémantique *LaCAS – Open Archive in Language and Cultural Area Studies*[8]. Comme nous l'expliquerons ci-après, LaCAS est une plateforme sémantique qui est destinée à collecter et à traiter les données de recherche provenant des études aréales, i.e. des recherches pluridisciplinaires dédiées à une zone géopolitique, linguistique, culturelle ou historique.

L'exemple illustre la distinction entre *donnée* et *métadonnées* dans le cadre d'une pratique d'analyse d'une ou d'un corpus de ressources documentaires :

1. la *partie 1* comprend la donnée documentaire (ici : l'enregistrement audiovisuel d'un chercheur qui parle de son objet de connaissance) ;
2. la *partie 2* comprend les différents éléments du métalangage de description qui sont mobilisés pour décrire la donnée.

La *partie 2* montre une variété de points de vue qui sont mobilisés pour expliciter l'enregistrement audiovisuel (*partie 1*). Chaque point de vue est représenté par un type particulier de métadonnées parmi lesquels nous trouvons celles que nous venons de mentionner. Soulignons que les différents points de vue identifiés dans la partie 2 de la capture d'écran (figure 2) reposent sur un *modèle de données*, ici, sur un *modèle de données textuelles* qui guide le travail d'analyse (d'expertise) documentaire. Un modèle de données (textuelles, dans notre cas) :

1. d'une part, explicite, une *certaine vision* de l'objet à analyser, à expertiser ;

---

[8] https://lacas.inalco.fr/

2. d'autre part, sert de *standard*, de *référence* aux activités d'expertise.

En d'autres termes, un modèle de données est une *structure d'ordre* et représente, dans ce sens, une référence culturelle – une *théorie*, un *référentiel métier* ou un *« benchmark »* (en français, *étalon*). Comme référence culturelle, le modèle de données est une *norme* et représente, en même temps, un *savoir* et *savoir-faire* (pour, dans notre cas, analyser, expertiser, une donnée documentaire).

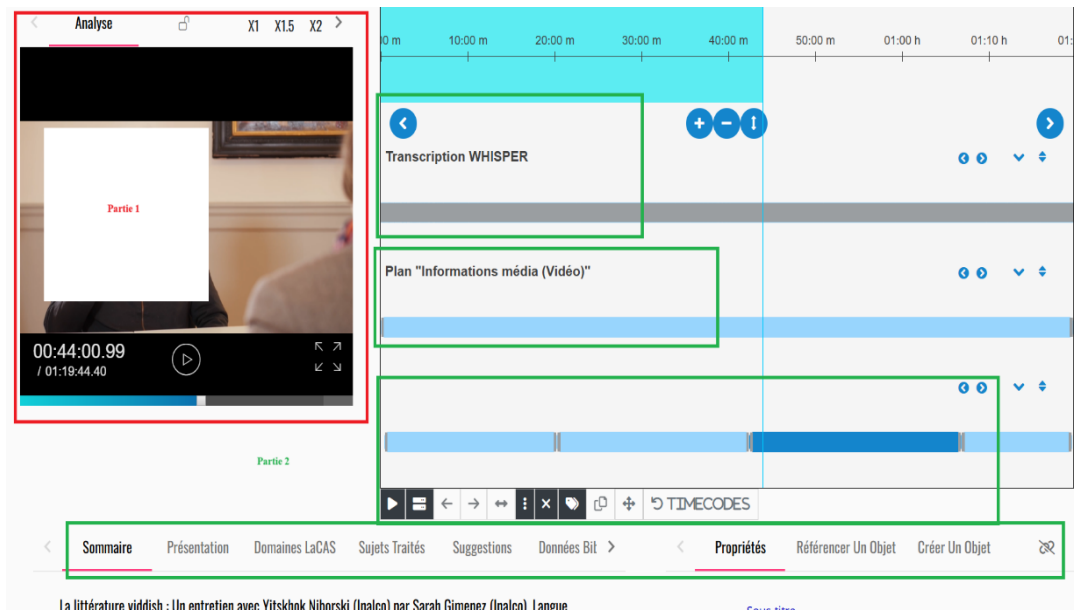

(Figure 2)

Cependant, soulignons que dans le cadre du Web sémantique, la distinction entre donnée et métadonnée *n'est plus liée à un type de données particulières*. Elle n'est plus liée à la « donnée documentaire » mais dépend tout simplement et d'une manière complètement générale du *point de vue* qu'on adopte pour affirmer qu'une *entité* est décrite (spécifiée…) par une ou un ensemble d'autres *entités* :

1. l'entité qui est décrite occupe le rôle de la *donnée ;*
2. l'entité ou les entités qui décrivent occupent le rôle des *métadonnées*.

## 4) EXEMPLE D'UNE PLATEFORME SEMANTIQUE DEDIEE A L'EXPERTISE AREALE

La notion *écosystème de base de connaissance* peut être compris comme un environnement sociotechnique *d'usage, de gestion* et *de gouvernance* d'une base de connaissance. C'est un *milieu artificiel* (i.e. un artefact, délibérément conçu et réalisé) pour collecter, traiter et exploiter de données de toutes sortes afin de produire et de faire circuler des savoirs ou savoir-faire en fonction d'une communauté d'acteurs et de leurs activités[9]. Il existe aujourd'hui beaucoup d'initiatives et de programmes qui offrent des outils et méthodes dans cette direction. En Europe, une des initiatives les plus importantes est l'EOSC, l'*European Open Science Cloud*[10].

Un écosystème à base de connaissance offre donc un milieu de travail dont l'environnement se compose typiquement de plusieurs *modules* qui prennent en charge la réalisation d'activités qui concourent à un *projet d'expertise*. Les deux principaux modules sont :

1) le module de l'exploration et de l'appropriation d'une base de connaissance par une population d'utilisateurs ;

2) le module de l'édition et de la gestion d'une base de connaissance par des acteurs habilités.

Ces deux modules peuvent se différencier, à leur tour, en des modules plus spécialisés. Ainsi, le module « édition et gestion » se différencie en :

- un module réservé à la *modélisation ontologique* (du ou des *domaines de connaissance*, des *données textuelles* lato sensu, des *activités de traitement* des données, du *contexte L.O.D.*) ;

- un module réservé à la *réalisation* et au suivi des *ressources métalinguistiques* (vocabulaire contrôlé, thésaurus…) ;

- un module réservé à la *réalisation* et au *suivi du portail sémantique* donnant accès à la base de connaissance et offrant des services à son appropriation et exploitation ;

- un module réservé au *traitement des données textuelles* (i.e. à la *curation* et au *moissonnage* des données textuelles sur le web ou dans des « entrepôts » de données en ligne, à l'*indexation* et à la *classification* des données textuelles, à leur *segmentation*, à leur *description* et *annotation*, à la mise en relation d'une donnée textuelle avec d'autres données textuelles…) ;

- un module réservé à la *définition des formats* de publication et à la *publication* à proprement parler d'un élément ou d'un ensemble d'éléments de la base de connaissance composée de données textuelles annotées, de collections (thématiques, auctoriales…) de données textuelles, de concepts documentés du thésaurus, de traductions ;

---

[9] Cf. également : Arduin, Pierre-Emmanuel, Grundstein, Michel, Rosenthal-Sabroux, Camille (2015)
[10] https://digital-strategy.ec.europa.eu/en/policies/open-science-cloud

- un module réservé à la *gestion* de l'écosystème de la base de connaissance (de la communauté de ses utilisateurs et de leurs droits, de la sauvegarde de la base de connaissance ; …).

Un écosystème de base de connaissance se déploie sur une infrastructure technologique souvent très complexe qui comprend notamment (mais pas exclusivement) un *environnement logiciel* permettant la planification, l'organisation et la réalisation des activités concrètes des divers projets menés en son sein.

Considérons maintenant un exemple concret d'un écosystème de base de connaissance. Il s'agit de la plateforme *LaCAS – Open Archive in Language and Cultural Area Studies* (figure 3). LaCAS est un projet que la vice-présidence scientifique de l'Inalco (cf. Stockinger 2023 ; Grossemy et al 2024) a réalisé entre 2020 et 2024 en partenariat avec plusieurs institutions françaises (Université de Paris Cité, Région Île-de-France, Ina…).

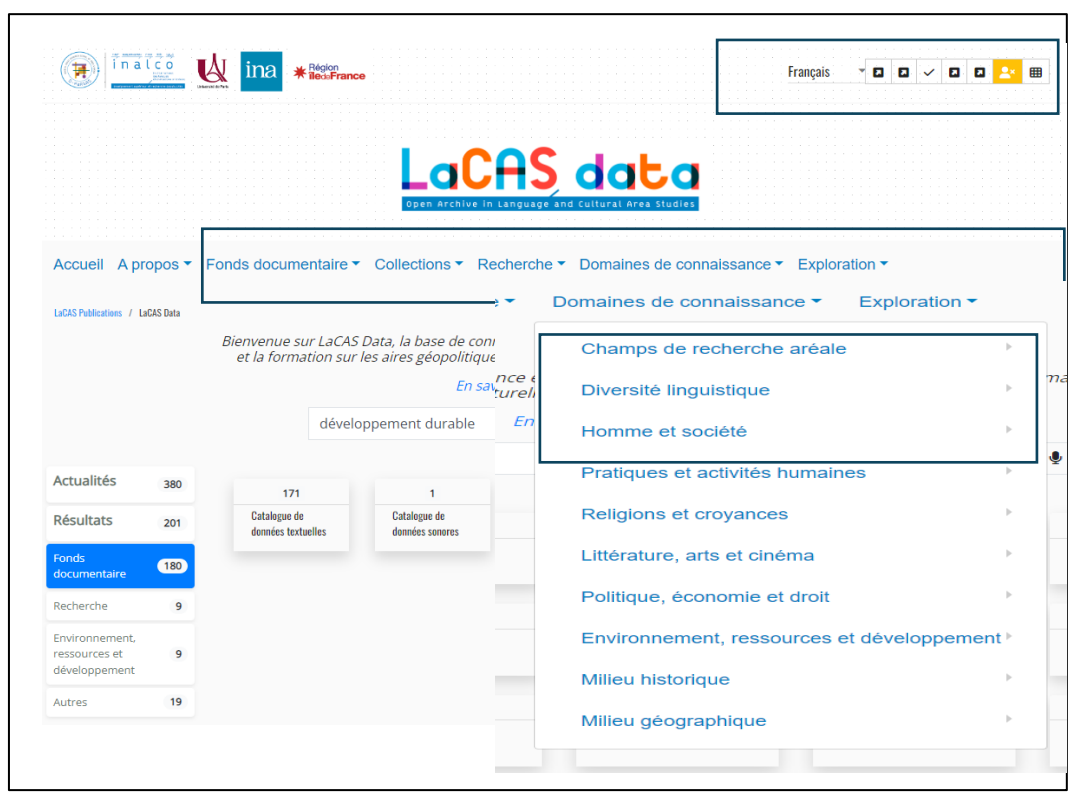

(Figure 3 : Page d'accueil du portail LaCAS data[11])

LaCAS est une plateforme qui se propose de collecter et de « traiter » (d' « expertiser ») les diverses *données de recherche* en études aréales. Par études aréales (ou régionales), on entend toutes les activités pluridisciplinaires dédiées à la connaissance d'une région donnée. C'est bien dans ce sens qu'on parle, par exemple, des études chinoises (ou de la sinologie), des études arabes, des études allemandes ou encore des études sahéliennes, des études baltiques, des études mongoles, voire des études coptes ou araméennes. Une « région », comme ces exemples le montrent, peut se manifester sous forme d'une entité politique, d'une zone géographique, d'une aire linguistique ou culturelle, d'une entité historique, etc. L'invariant qui est commun à ces études est qu'elles représentent des efforts mono- ou pluridisciplinaires pour mieux appréhender – dans une perspective critique et comparative - une région particulière, délimitée dans l'espace et le temps, tantôt dans sa totalité, tantôt par le biais d'une de ses caractéristiques culturelles, langagières, sociales, politiques ou autres.

---

[11] https://urlr.me/BZNWpz

La plateforme LaCAS est constituée autour d'une *base de connaissance* (= *base de connaissance LaCAS*) comprenant un *thésaurus* spécialisé dans les études aréales (= *thésaurus LaCAS*) et d'un fonds documentaire qui est moissonné dans divers entrepôts de données scientifiques en ligne - principalement dans HAL du CNRS[12] ainsi que, dans une moindre mesure, dans Nakala (CNRS également)[13] et Zenodo (CERN)[14] (= *fonds documentaire LaCAS*). Le fonds documentaire LaCAS comprend actuellement (juillet 2025) environ 160.000 données textuelles stricto sensu (publications, rapports, thèses…), visuelles, audiovisuelles et sonores. Il s'enrichit quotidiennement à l'aide d'un système de *moissonnage automatique* (en ce qui concerne HAL) et de *curation-moissonnage supervisé* par un expert humain pour les autres entrepôts (Nakala et Zenodo, pour le moment).

Comme on peut le voir dans la figure 3, la base de connaissance LaCAS regroupe en 10 grands macro-domaines l'énorme diversité des objets (appelés *« objets de connaissance »*) en études aréales qui sont documentés par les données et travaux de recherche des chercheurs en études aréales et qui font partie du fonds documentaire LaCAS. Chaque macro-domaine de connaissance se différencie en un ou plusieurs domaines de connaissance plus spécialisés. Un objet de connaissance peut appartenir à un ou plusieurs domaines de connaissance et est qualifié par un *schéma conceptuel* (i.e. par un graphe ou une configuration de propriétés). Selon le cas, un schéma conceptuel qualifiant la structure d'un objet de connaissance peut être plus ou moins complexe ou élémentaire. Nous discuterons un exemple concret dans le chapitre 6.

Le schéma (ou modèle) conceptuel sert de référence ou de *standard* pour toutes les activités qui traitent un objet de connaissance. Une activité tout à fait centrale est celle de la *documentation* d'un objet de connaissance tel que, par exemple, d'une langue particulière, d'un courant littéraire, de la politique économique d'un pays, d'un secteur industriel d'une région géopolitique, des relations diplomatiques entre deux pays, d'un processus historique, d'une écorégion du monde, etc. La documentation repose sur une *expertise* sous forme :

1) d'*alignements sémantiques* de l'objet de connaissance sur les grands référentiels du L.O.D. (Wikidata, Bnf Rameau, Idref, Eurovoc, Loterre…) ;

2) de *présentations* rédigées « manuellement » (i.e. rédigées par un acteur habilité) ;

3) de *corpus ouverts* de données de recherche ;

4) d'une sélection de *références webographiques* réalisée à l'aide de moteurs spécialisés tels qu'Isidore du Cnrs, OpenAlex, BnF Gallica ou Semantic Scholar ;

5) d'*informations institutionnelles* (équipes de recherche, projets de recherche, chercheurs).

L'expertise comprend également des « éclairages » générés d'une manière (semi-)automatique à l'aide d'outils appelés « agents conversationnels » provenant de l'IA neuronale générative. Afin de procéder à la localisation et à l'analyse automatique d'un nombre quantitativement plus ou moins élevé de données disponibles, la génération de ces expertises à l'aide d'« agents conversationnels » s'appuie sur l'ontologie LaCAS qui impose, comme déjà dit, un schéma conceptuel à chaque objet de connaissance. Nous reviendrons sur cette nouvelle pratique de production d'une expertise « guidée par l'IA » un peu plus loin (voir chapitre 7).

---

[12] https://hal.campus-aar.fr/AAI
[13] https://nakala.fr/collection/10.34847/nkl.cc43rwm3
[14] https://zenodo.org/

La base de connaissance LaCAS, les domaines et les objets de connaissance qu'elle abrite ainsi que le fonds documentaire LaCAS sont accessibles via deux portails web : *LaCAS data*[15] et *LaCAS publications*[16]. LaCAS data offre des services pour interroger la base de connaissance, consulter la documentation d'un objet de connaissance, explorer le fonds documentaire LaCAS et ses diverses collections. LaCAS publications, de son côté, offre des collections de republications de données et de métadonnées LaCAS sous forme de « dossiers dynamiques » dédiés à une exploration très structurée et guidée d'une aire culturelle ou géopolitique donnée et/ou d'une problématique particulière en études aréales.

Le fonctionnement de l'environnement de la plateforme est assuré par un environnement logiciel, développé par l'équipe de recherche de l'Ina (Institut national de l'audiovisuel) à Paris. L'acronyme de cet environnement est *Okapi* (*Open Knowledge Annotation and Publishing Interface*)[17] et est documenté dans plusieurs publications (Beloued et al 2025, 2015 ; Lalande et al, 2022). L'environnement logiciel Okapi est déployé sur une infrastructure informatique propre à l'Inalco (Institut national des langues et Civilisations Orientales).

Un écosystème de base de connaissance tel que celui incarné par la plateforme LaCAS rend possible et, simultanément, contraint la conception et l'exercice d'une grande variété d'activités de recherche (dans notre cas : en études aréales). Des exemples de *projets LaCAS* incluent la constitution d'archives numériques de recherche sur une aire particulière ; l'indexation, la classification sémantique et l'annotation de corpus de données ; la contextualisation ou la recontextualisation (auctoriale, institutionnelle…) de données non-qualifiées ou dont la qualification pose un problème ; la production de cartographies de la recherche sur de domaines et/ou d'objets de connaissance particuliers (par exemple, sur une aire linguistique ou sur les politiques de santé dans une région du monde, etc.).

Les diverses activités qui marquent la « vie » dans un écosystème tel qu'il s'incarne dans la plateforme LaCAS se répartissent en plusieurs grands « secteurs » que sont :

1. la *modélisation conceptuelle* d'un domaine d'expertise qui constitue le domaine de référence d'une base de connaissance. Les activités de modélisation comprennent plus particulièrement le design et la réalisation de l'ontologie du domaine d'expertise.

2. la *conception*, la *réalisation* et le *suivi* d'un *thésaurus* ou d'un *vocabulaire contrôlé*. Les termes d'un thésaurus désignent un objet de connaissance particulier relevant d'un ou de plusieurs domaines de connaissance. Ils remplissent deux fonctions centrales : d'une part, ils servent comme *outil d'indexation* et de *classification* de données textuelles lato sensu et ils constituent, d'autre part, des *agrégateurs* (des « *hubs* ») de données et d'informations sur un objet de connaissance.

2. la *localisation*, la *curation*, le *moissonnage* et l'*import* de corpus de données textuelles dans une base de connaissance. Une base de connaissance « vivantes » s'enrichit continuellement de nouvelles données textuelles au sens large du terme.

---

« Données textuelles au sens large du terme » signifie simplement tout élément qui apporte une (nouvelle) information à une base donnée. Cet apport se fait sous forme d'un projet qui mobilise typiquement les actions mentionnées de localisation d'une donnée (par exemple, sur le web, dans un entrepôt de données ou encore sous forme d'une production ciblée de données manquantes), de *curation* d'une donnée (sous forme d'une évaluation-sélection en fonction de sa conformité aux critères de qualité et au modèle organisant la base de connaissance) ainsi que de moissonnage et d'import d'une donnée (sous forme, par exemple, de la transcription de certaines de ses métadonnées – auctoriales, descriptives… - et de son insertion dans la base de connaissance).

3. l'*analyse* des données ou des corpus de données textuelles. L'analyse peut consister en des activités plus ou moins répétitives et automatisées du genre indexation de données. Cependant, elle peut également prendre la forme de projets complexes de recherche et d'expertise consacrés à un objet de connaissance particulier. Ainsi, l'analyse s'incarne dans une diversité d'activités telles que le regroupement de données pour constituer un corpus de travail, la segmentation de données complexes pour mieux extraire des informations pertinentes, la description critique et la comparaison de données, la classification et la cartographie des données ou encore la synthèse analytique qui conclut un projet.

4. la *réutilisation de données*. La réutilisation d'une donnée (« originale », i.e. telle qu'elle est enregistrée la première fois dans une base de connaissance) est comprise ici au sens large du terme. Elle inclut tout projet de production de nouvelles *versions* (modifiées) d'une donnée originale, de *montage* et de *fusion* de données ou encore de *publication* de données originales et/ou modifiées. La réutilisation de données joue un rôle central dans tout projet de publication, de valorisation, de médiation ou encore de réemploi d'une donnée dans des contextes d'usage autres que celui pour lequel la donnée « originale » a été conçue et réalisée.

5. La gestion de l'écosystème et de ses différents constituants et réalisations. La gestion de l'écosystème comprend celle des acteurs qui constituent sa communauté des parties prenantes, celle du suivi de la base de connaissance et de ses divers composants, celle des portails de l'écosystème et des diverses interfaces et, enfin, celle du logiciel lui-même et de son déploiement sur une infrastructure appropriée.

En considérant les diverses activités citées ci-dessus, on peut aisément se convaincre de la place centrale de la notion de *cycle de vie* d'une donnée textuelle (lato sensu) qui documente un objet de connaissance (en études aréales). Le cycle de vie d'une donnée textuelle est intimement conditionné par les diverses activités de *traitement* dont une donnée textuelle (ou un corpus de données textuelles) fait l'objet. Ces activités font partie et s'intègrent dans un processus que nous avons appelé *expertise*.

L'expertise peut être comprise dans un sens plus restreint, comprenant les divers efforts d'*analyse* et d'interprétation d'une donnée ou, plutôt, d'un corpus de données. Elle peut être comprise également dans un sens plus large, y compris aussi bien les diverses activités « en amont » et « en aval » d'une analyse. « En amont » de l'analyse d'une donnée, on trouve notamment les activités de *localisation* (dans un entrepôt de données), de *curation*, de *moissonnage* et d'*import* d'une ou d'un corpus de données dans une base de connaissance. En aval de l'analyse d'une donnée, on trouve les activités éditoriales pour produire des « nouvelles » données textuelles lato sensu à partir d'un corpus de données déjà analysées. Ce processus d'expertise au sens large du terme est problématisé et décrit plus en détail dans (Stockinger 2012).

L'expertise peut servir tantôt à l'enrichissement de la base de connaissance elle-même, tantôt à la satisfaction de besoins relevant de pratiques et institutions qui sont parties prenantes de l'écosystème d'une base de connaissance, mais qui doivent répondre à des enjeux et objectifs qui sont les leurs. Dans le premier cas, l'expertise en matière de données sert à améliorer, à enrichir, à diversifier et aussi à tester la solidité conceptuelle et la qualité d'une base de connaissance existante ; dans le second cas, l'expertise « exploite » une base de connaissance afin de répondre à des objectifs de plus en plus variés, qu'ils soient professionnels, institutionnels, politiques, économiques ou autres, selon le cas. Cela dit, ces deux orientations, qui conditionnent l'activité d'expertise en matière de données, ne s'excluent pas mutuellement ; elles sont complémentaires et se renforcent mutuellement.

## 5) LA BASE DE CONNAISSANCE LACAS

Une base de connaissance telle que celle dont est pourvue la plateforme LaCAS comprend une grande diversité d'*objets de connaissance* qui font partie d'un ou de plusieurs *domaines de connaissance*.

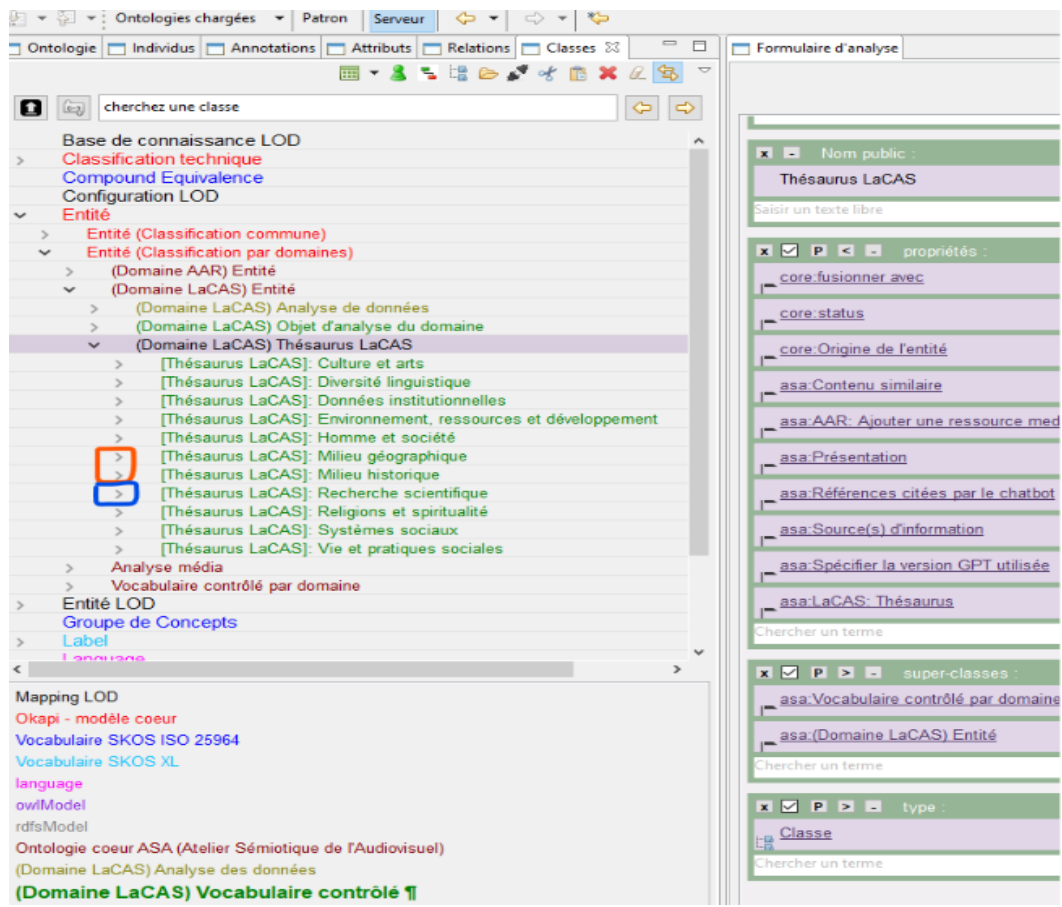

(Figure 4)

Comme déjà dit, Le domaine d'expertise en études aréales que couvre la base de connaissance LaCAS est organisé en dix macro-domaines (figure 3). On y trouve, entre autres, les macro-domaines de connaissance « Diversité linguistique », « Homme et société », « Vie et pratiques sociales », « Religions et spiritualité », « Culture et arts » et « Systèmes sociaux ». Un macro-domaine de connaissance à son tour se différencie en des domaines de connaissance plus circonscrits. Ainsi, par exemple, le macro-domaine « Diversité linguistique » se différencie en « Langues du monde », « Groupes et familles de langues », « Aires linguistiques », « Recherche sur les langues », « Traduction et interprétation », « Traitement automatique des langues » ou encore « Apprentissage et enseignement des langues ». Il va sans dire qu'un macro-domaine de connaissance tel que celui intitulé « Diversité linguistique » peut être enrichi par d'autres domaines. La catégorisation des objets de recherche en macro-domaines et domaines de connaissance plus spécialisés est le résultat de décisions du *design conceptuel* de la base de connaissance dont les justifications sont, en règle générale, de nature empirique et/ou institutionnelle.

La figure 4 montre un petit extrait de l'*ontologie du domaine LaCAS* qui comprend les principales *classes* appelées [(*Domaine LaCAS*) *Thésaurus LaCAS*]. Le choix de ces classes et leur dénomination sont toujours des décisions conceptuelles et terminologiques délicates. Soulignons cependant qu'il est établi sur la base de trois *grands axes sémantiques*, à savoir :

1. l'axe *approches, disciplines, méthodes, problématiques* pour étudier une aire ;

2. l'axe *objets de recherche* qui font partie d'une aire ;

3. l'axe *contexte géographique et historique* qui situe et délimite une aire dans l'espace et le temps.

Ainsi, la classe [Recherche scientifique] est réservée aux aspects théoriques, méthodologiques, disciplinaires, etc. Les deux classes [Milieu géographique] et [Milieu historique] sont réservées au contexte géographique et historique qui situe et délimite un objet de connaissance étudié. Toutes les autres classes qui dépendent de la super-classe [(Domaine LaCAS) Thésaurus LaCAS] (et à la seule exception de la classe [Données institutionnelles]) identifient et catégorisent les multiples objets de connaissance qui font partie du domaine d'expertise en études aréales de LaCAS.

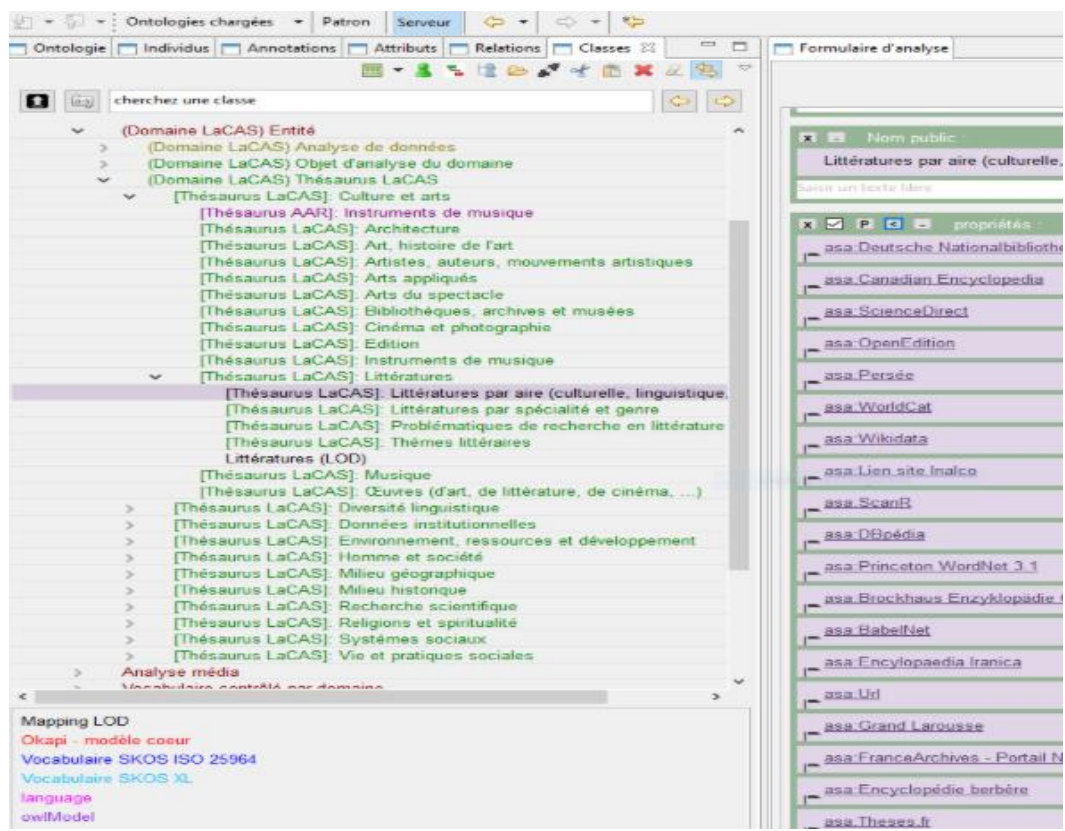

(Figure 5)

En comparant les classes identifiées dans la figure 4 avec les 10 macro-domaines de connaissance du thésaurus LaCAS dans la figure 3, nous pouvons constater certaines différences entre « classes » et « domaines ». Ces différences s'expliquent par le fait que les classes font partie du *modèle conceptuel* – de l'ontologie – *du domaine* de référence

LaCAS, tandis que les macro-domaines et les domaines de connaissance appartiennent au *modèle conceptuel* – à l'ontologie – *de la visualisation* (ou de la *publication*) des classes. Nous ne pouvons pas entrer ici dans les détails, mais précisons simplement qu'une base de connaissance (qui est régie par un modèle conceptuel ou une ontologie de domaine) peut être visualisée de nombreuses façons différentes. La visualisation d'une base de connaissance et de ses données dépend, en effet, d'une « stratégie » de *médiation* et de *médiatisation*, dont la réalisation est conditionnée par l'existence d'un modèle (ou d'un « scénario ») définissant cette stratégie.

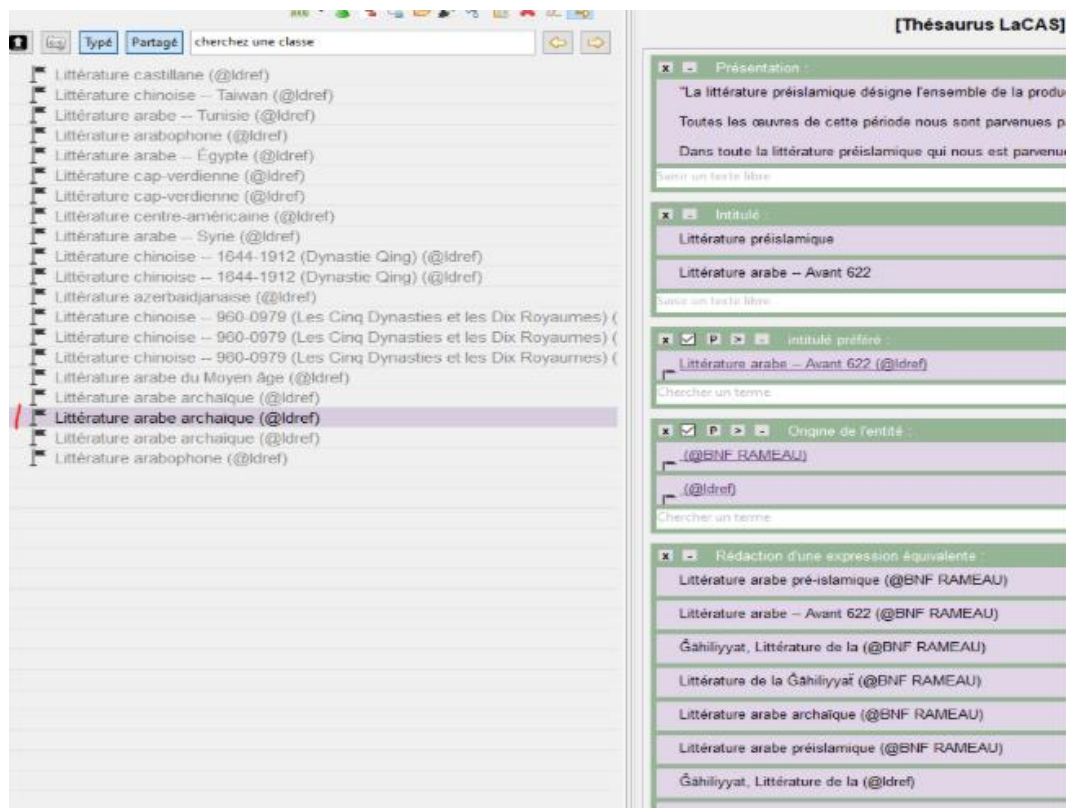

(Figure 6)

Comme le montre la figure 5, une classe sémantique peut se différencier en une hiérarchie partielle (ou *treillis*) de sous-classes plus spécialisées. La figure 5 nous montre l'exemple de la classe [Littératures] qui fait partie de la super-classe [Culture et arts]. La classe [Littératures] se différencie en des objets de connaissance qui font partie tantôt de la classe [Littératures par aires (linguistiques, géographiques…)], tantôt de la classe [Littérature par spécialité et genre], tantôt de la classe [Problématique de recherche littéraire], tantôt encore de la classe [Thèmes ou motifs littéraires]. Ainsi, par exemple, l'objet de connaissance « Littérature romanesque albanaise » fait partie des deux premières classes ([Littératures par aires], [Littérature par spécialité et genre]), l'objet de connaissance « Techniques de narration – roman japonais contemporain » fait partie à la fois de ces deux classes mentionnées ainsi que de la classe « [Problématique de recherche littéraire] ; enfin l'objet de connaissace « Thème (topos) de la jeunesse -- Littérature maghrébine » fait partie des deux classes [Thèmes et motifs littéraires] et [Littératures par aires (linguistiques, géographiques…)].

Bien entendu, d'autres classes peuvent être ajoutées aux classes sémantiques existantes et une classe donnée peut être à son tour différenciée en une ou plusieurs classes plus spécialisées. En revanche, il est plus délicat de redéfinir une classe, par exemple, en la scindant en deux ou plusieurs classes, en la fusionnant avec d'autres classes ou en la supprimant.

Une classe sémantique contient entre 1 à n objets de connaissance. Techniquement, on parle, au lieu d'objet de connaissance, plutôt d'*individu* ou d'*entité*. Par exemple, la classe [Littératures par aires (linguistiques, géographiques, …)] comprend actuellement (juillet 2025) quelque 230 objets de connaissance (i.e. *individus* ou *entités*) parmi lesquels nous trouvons l'objet « Littérature arabe – Avant 622 », appelée également « Littérature arabe archaïque » par le référentiel *Idref*[18] (qui s'aligne, entre autres, sur la terminologie RAMEAU de la Bibliothèque nationale de France). La figure 6 nous fournit un petit extrait montrant comment cet objet de connaissance est présenté dans le thésaurus LaCAS.

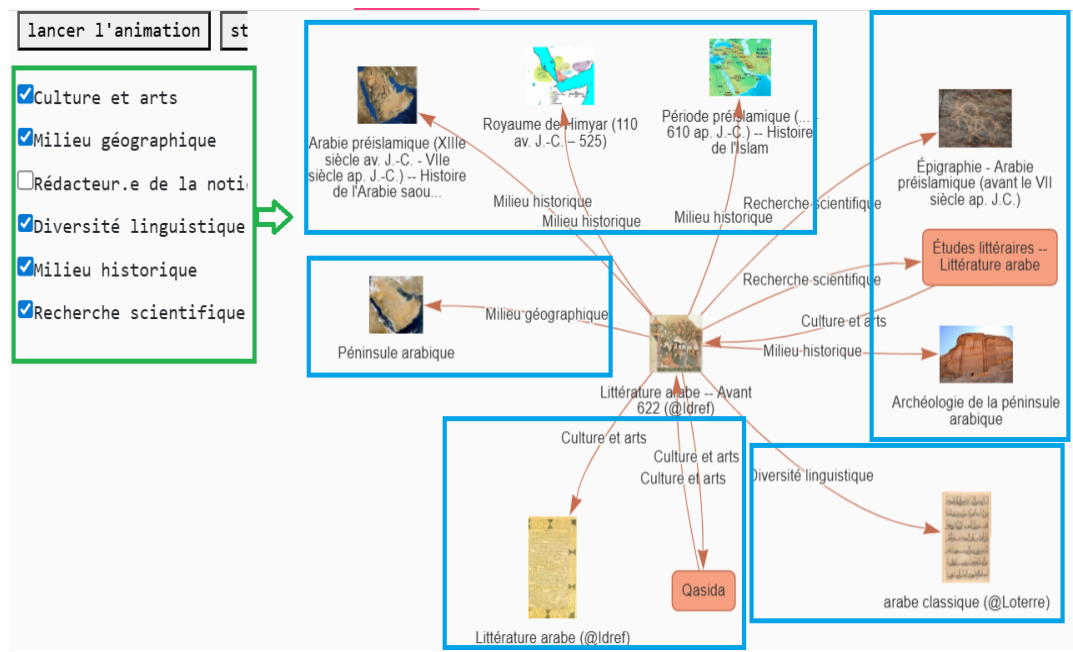

(Figure 7)

Ainsi, le thésaurus LaCAS comprend actuellement (juillet 2025) quelque 8500 objets de connaissance parmi lesquels nous comptons les objets « Littérature arabe – Avant 622 » et le « Quechua » que nous discuterons plus en avant dans le prochain chapitre. Chaque objet est décrit et documenté avec plus ou moins de finesse. Certains objets sont plus élaborés que d'autres ce qui est une conséquence de divers biais – *biais institutionnels* (intérêt d'une institution de recherche pour tel ou tel objet de connaissance), *biais liés à l'existence* même de données documentant un objet ; *biais liés à l'origine* et à la qualité de données ; biais, enfin, liés à la *subjectivité inhérente* à la production d'une expertise.

Cela nous amène à ajouter encore un mot au sujet de la *grille de description* utilisé pour produire une documentation structurée des objets appartenant à un domaine de connaissance (à une même classe sémantique). Cette grille fait partie du modèle conceptuel du domaine d'expertise LaCAS et comprend notamment les *propriétés* que partagent les objets de connaissance d'une même classe. Comme on peut le voir dans la figure 5, les propriétés d'une classe sont identifiées dans la partie médiane de l'interface de l'éditeur d'ontologies Okapi. De nouveau, sans rentrer dans trop de détails, certaines propriétés possèdent une portée *très générale* (tous les objets de la base de connaissance en sont pourvus), d'autres sont tout à fait *spécifiques* aux objets de connaissance d'une classe. Par ailleurs, chaque propriété est pourvue d'un certain « comportement fonctionnel » qu'il faut définir dans l'éditeur d'ontologies Okapi (cf. ci-après la figure 10).

---

[18] https://www.idref.fr/

Parmi les propriétés les plus récurrentes, on trouve la « désignation (de l'objet de connaissance) », la « présentation (de l'objet) », les « informations (sur l'objet) produites par une I.A. neuronale générative » ou encore le « positionnement de l'objet par rapport aux autres objets de connaissance LaCAS ». Pour illustrer cette dernière propriété, considérons l'objet « Littérature arabe – Avant 622 ». Comme nous le montre la figure 7, cet objet est connecté à d'autres objets de connaissance qui précisent le contexte géographique (« Péninsule arabique »), les époques et formations historiques concernées (« Période préislamique – Histoire de l'Islam », « Royaume de Himyar » ; …), les disciplines et approches scientifiques concernées (« Archéologie de la péninsule arabique », « Épigraphie – Arabie préislamique », « Études littéraires – Littérature arabe »), les langues concernées (« Arabe archaïque », « Langues sud-arabiques » ; …), les littératures concernées (« Littérature arabo-musulmane »), les genres et courants littéraires concernés (« Poésie préislamique », « Qasida » ; …).

Les objets ainsi connectés forme des *configurations de connaissance*, typiquement représentées sous forme de *graphes* de plus en plus complexes. C'est dans ce sens qu'on dit qu'une base de connaissance telle que celle de la plateforme LaCAS forme un énorme graphe de connaissance, dense et toujours en évolution, dans lequel chaque objet se trouve contextualisé dans une *configuration locale* caractéristique. Ainsi, une configuration (locale) de connaissance telle que celle dans laquelle on trouve l'objet « Littérature arabe – Avant 622 », peut évoluer dans le temps. Il peut s'enrichir de nouveaux objets, se différencier en des configurations plus locales, etc.

## 6) EXEMPLE : PRODUIRE UNE DOCUMENTATION STRUCTUREE SUR UNE LANGUE

Le domaine de connaissance « Langues du monde » comprend actuellement (i.e. en juillet 2025) une documentation de quelque 540 langues.

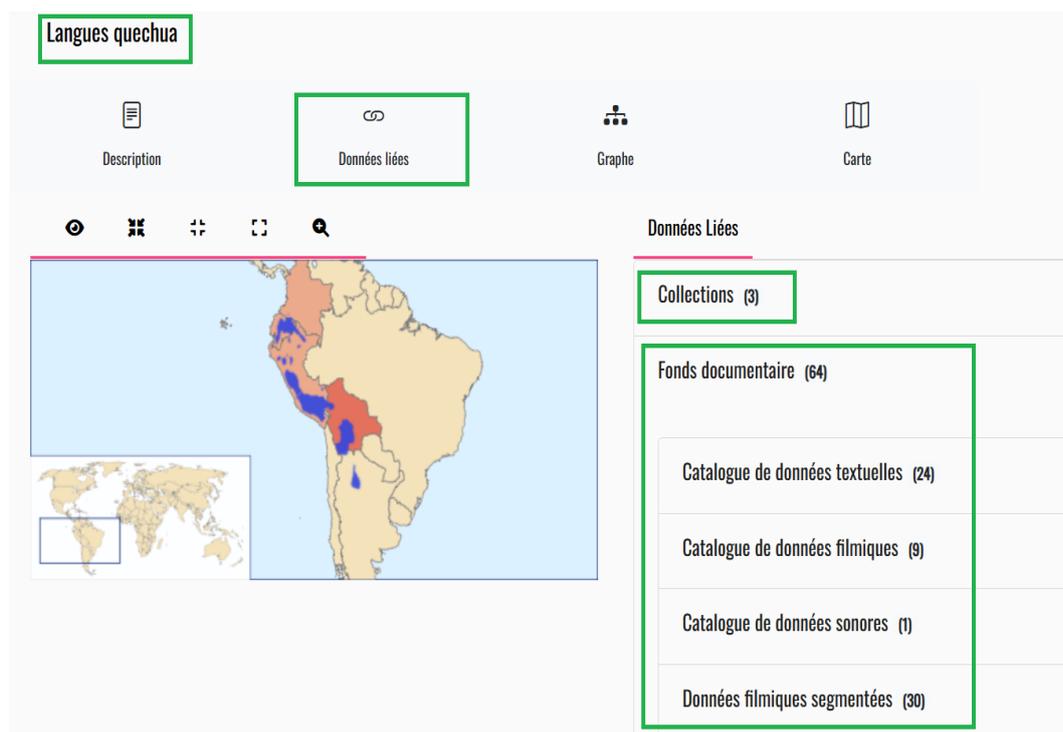

(Figure 8)

Pour en fournir, comme expliqué dans le chapitre précédent (chapitre 5), une expertise sous forme d'une *documentation structurée*, il nous faut d'abord un *modèle conceptuel* qui sert de référence au travail d'*expertise* réalisé soit par un ou une communauté d'acteurs humains (chercheurs, cogniticiens, etc.), soit grâce à des méthodes et des outils numériques et/ou basés sur l'IA neuronale pour la localisation, la qualification et l'analyse des données (textuelles au sens large).

Prenons l'objet de connaissance « Langues quechua », le parler vernaculaire d'une importante population de locuteurs principalement localisée dans les Andes. La figure 8 montre la capture d'une partie du *hub de connaissances et d'informations* qui est dédié à cet objet de connaissance. Il identifie notamment les quelques 64 ressources documentaires identifiées et moissonnées dans les entrepôts HAL, Zenodo et Nakala. Ces ressources documentaires comprennent, entre autres, des publications, rapports scientifiques, cours, documentations visuelles, enregistrements audiovisuels, extraits filmiques.

La figure 9 nous montre un extrait de la *visualisation* du graphe de connaissance de notre hub. Cette visualisation nous apprend que l'objet de connaissance « Langues quechua » peut être appréhendé par un ensemble de caractéristiques ou propriétés que se partagent en effet tous les objets de connaissance (i.e. toutes les entités) qui relèvent de la *classe sémantique* de l'ontologie LaCAS « Langues du monde ».

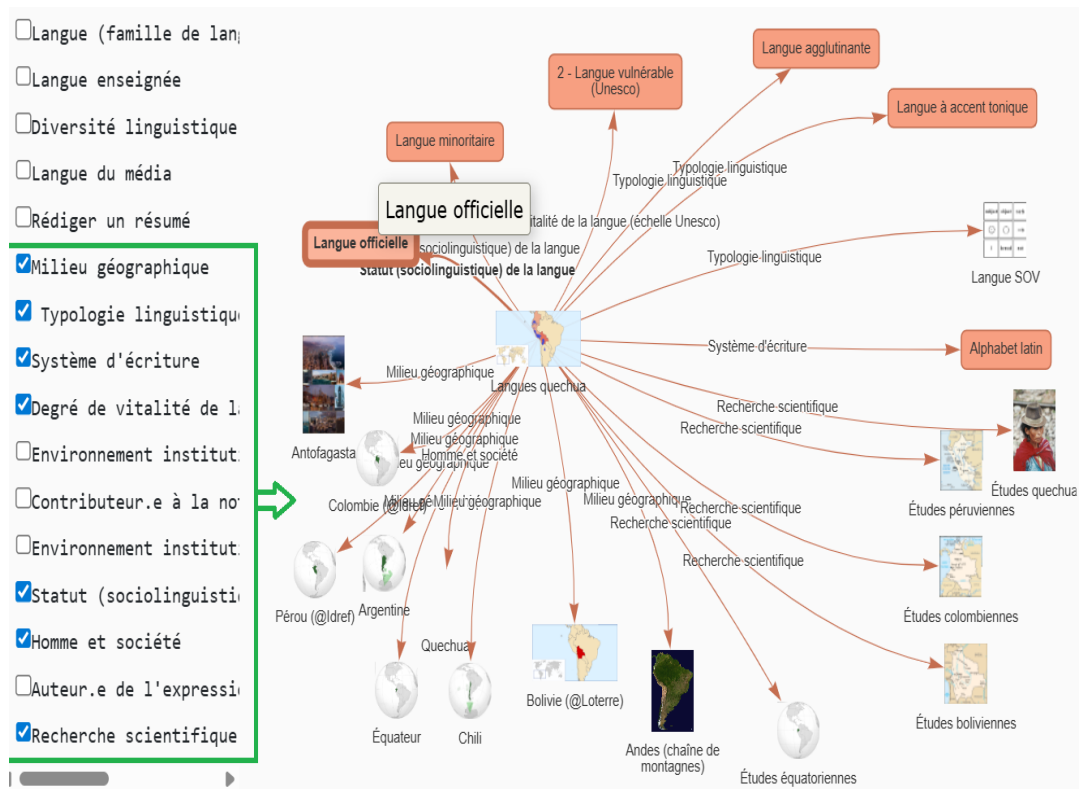

(Figure 9)

Certaines caractéristiques relèvent d'une propriété appelée « Typologie linguistique », d'autres dépendent d'un propriété appelée « Milieu géographique » et d'autres encore font partie des propriétés appelées « Statut sociolinguistique », « Homme et société », « Degré de vitalité de la langue », etc. Une propriété peut se différencier en un ensemble de sous-propriétés (de propriétés plus spécialisées). Par exemple, la propriété « Typologie linguistique » comprend une variété de propriétés qui permettent de décrire de manière fort détaillée les caractéristiques phonologiques, morphosyntaxiques et lexicales d'une langue. Comme le montre la figure 10, le « comportement fonctionnel » d'une propriété est conditionné par sa source et sa cible. Les exemples que nous fournit la figure 10, la source et la cible sont des classes sémantiques qui font partie de l'ontologie du domaine LaCAS. Ces propriétés sont appelées relations. A côté des relations, il existe encore des propriétés appelées « attributs » qui s'étendent entre une source qui est une classe) et la valeur littérale qui caractérise l'objet-source (une valeur littérale peut être un texte, un nombre, une date, une valeur booléenne, etc.). Das la figure 10, nous voyons spécifié une propriété dont la structure est très simple qui s'intitule « Spécifier le degré de vitalité de la langue (échelle Unesco) ». Il s'agit ici d'une *relation* qui spécifie (i.e. qui *énonce*, qui *affirme*) qu'il existe une relation entre tous les objets de la classe sémantique « Langues » et la classe sémantique « Degré de vitalité de la langue (échelle Unesco) » de sorte que chaque langue peut être caractérisée par au moins une, sinon plusieurs valeurs qui définissent l'échelle de vitalité établie par l'Unesco.

Ensemble, les propriétés (*relations* et *attributs*) définissent le *modèle conceptuel* servant de référence – de standard – à la production d'une expertise sur une langue sous forme d'une *documentation structurée*. L'expertise peut être conçue comme une activité unique ou comme un observatoire permanent qui actualise la documentation à mesure que de nouvelles informations sur la langue sont identifiées et qualifiées. Les caractéristiques (typologiques, sociolinguistiques, en termes de vitalité linguistique, géographiques, démographiques, etc.) d'une langue telle que le quechua consistent soit en une *description verbale* (ou autre) *structurée* (i.e. produite selon un modèle du résultat attendu de la documentation), soit en l'identification d'un *jeu de valeurs prédéfinies* dans lequel une propriété peut se réaliser (qu'elle soit ou non accompagnée d'une description structurée - verbale ou autre).

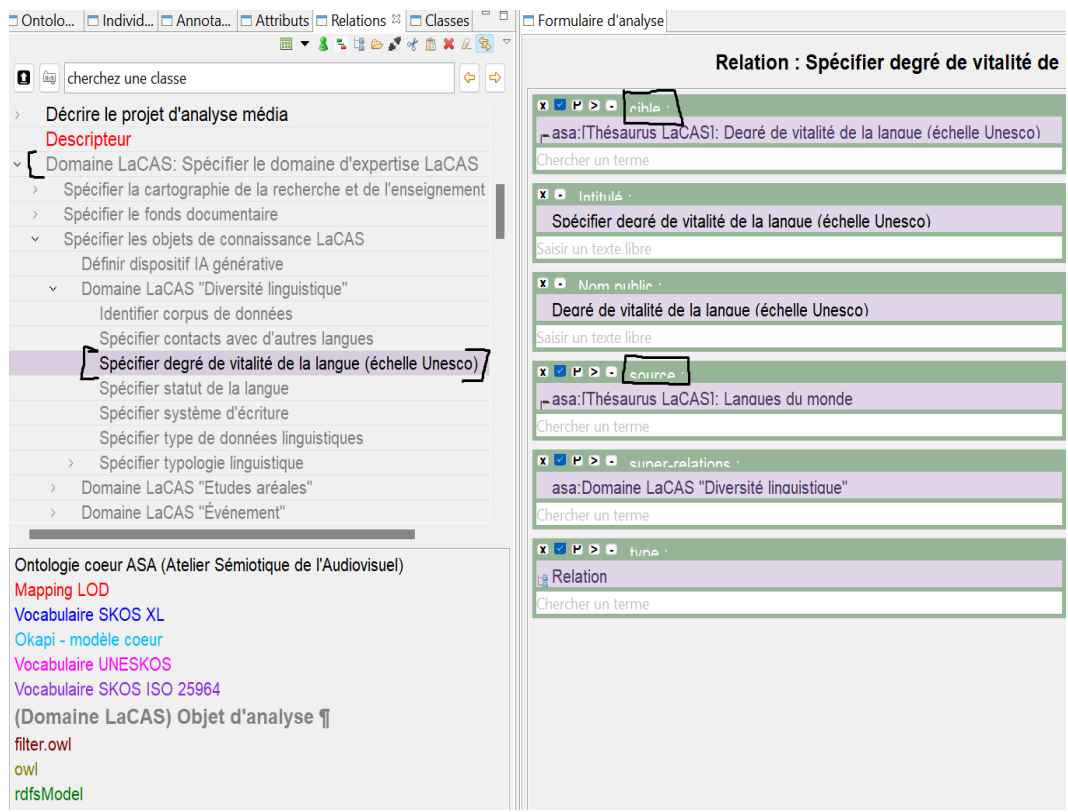

(Figure 10)

Ainsi, la figure 9 nous montre, plus précisément les caractéristiques qui qualifient l'objet de connaissance « Langues quechua » et le rapprochent ou l'éloignent d'autres objets de connaissance de la *même* classe sémantique. On trouve, par exemple, les caractéristiques « Langue agglutinante », « Langue à accent tonique », « Langue SOV », « Alphabet latin », « Langue vulnérable (selon l'UNESCO) », « Langue minoritaire », « Quechua (peuple) », « Pérou », « Bolivie », etc.

En d'autres termes, ce sont des *énoncés* qui affirment que le quechua est une *langue agglutinante* avec un *accent tonique*, que sa structure phrastique suit le modèle *sujet-objet-verbe*, que son statut sociolinguistique est considéré, selon l'échelle établie par l'UNESCO, comme *vulnérable*, qu'elle est la langue maternelle de la *population quechua* et qu'elle est parlée, entre autres, au *Pérou* et en *Bolivie*. Bien entendu, rien n'empêche de compléter cette liste d'énoncés prédiquant un ensemble de caractéristiques dont est pourvu le quechua. On pourrait s'imaginer des ajouts qui explicitent davantage les aspects relatifs aux usages sociolinguistiques d'une langue (par exemple, dans les médias, dans la vie politique, dans l'éducation…) ou aussi les scénarios de son évolution en contact avec d'autres langues. Tout dépend, bien entendu, du ou des *objectifs* que poursuit un projet de réalisation d'une base de connaissance documentant, pour rester avec notre exemple, les langues du monde et nos connaissances de ce patrimoine humain.

## 7) VERS UNE APPROCHE HYBRIDE « I.A. SYMBOLIQUE » - « I.A. NEURONALE »

L'avènement des méthodes de collecte et de classification de données se basant sur les avancées de l'apprentissage automatique (du « machine learning », en anglais) et son intégration dans une approche symbolique de modélisation explicite d'un domaine de connaissance ouvre des perspectives tout à fait nouvelles, prometteuses mais néanmoins exigeant une posture de prudence théorique et épistémologique. Ainsi, par exemple, les deux problèmes récurrents et centraux que rencontre tout projet de réalisation d'une base de connaissance sont :

1. celui du *choix des propriétés à retenir dans le design conceptuel du modèle du domaine* qui sert de standard aux activités d'expertise d'un ensemble d'objets de connaissance ;
2. celui de la *localisation, qualification et analyse à proprement parler de grands volumes de données pertinentes* pour documenter un objet de connaissance.

Les approches et méthodes en apprentissage automatique peuvent à priori contribuer à une bien meilleure prise en charge de ces deux problèmes. On peut ainsi s'attendre qu'elles sont capables de comparer les propriétés définissant un modèle conceptuel avec celles qui émergent, généralement sous forme de régularités statistiques et probabilistes, dans des très grands corpus de données textuelles (au sens large du terme). De même, la documentation empiriquement représentative d'un objet de connaissance (tel qu'une langue, par exemple) en référence à un modèle conceptuel riche en propriétés sophistiquées peut être conçue comme un processus d'expertise réalisé avec l'aide systématique de méthodes et d'outils issus de l'apprentissage automatique.

Prenons l'exemple de la figure 9 ci-dessus. La recherche et la classification correcte des informations morphosyntaxiques et/ou sociolinguistiques relatives à la langue quechua peuvent déjà bénéficier grandement des méthodes et des outils de l'*intelligence artificielle générative*. Les outils tels que les *agents conversationnels* sont en effet capables de localiser et d'extraire ces informations dans des larges *corpus* de données primaires et secondaires afin de les ajouter à celles produites « manuellement » par les spécialistes humains de cette langue. En prenant comme *entrée* les propriétés référencées dans la production d'une expertise sur cette langue, des *prompts spécialisés* contraignent et encadrent le travail (et le « raisonnement ») des agents conversationnels (des « chatbots ») qui utilisent un modèle de langage (LLM) tantôt *général* (tel que ChatGPT, Mistral ou DeepSeek), tantôt *spécialisé* (i.e. entraîné sur un corpus de données particulier). Ce travail duplique celui de l'expertise « humaine » et consiste, en gros, en la localisation correctement des sources d'information (répondant à un ensemble de critères de qualité), l'extraction des informations pertinentes, la comparaison et l'évaluation des informations extraites (conformément au modèle conceptuel qui sert de standard) et enfin en la génération de présentations verbales et/ou infographiques pouvant régulièrement être mises à jour.

Il s'agit de perspectives de recherche et de développement très récentes, dont il est difficile aujourd'hui de prédire les conséquences à moyen et long terme pour la recherche en sciences humaines et sociales en général, et pour les études aréales en particulier. Cela dit, dans le cadre de ce que nous venons d'esquisser concernant l'expertise en matière de données pertinentes pour l'étude d'une aire géopolitique ou culturelle, un référentiel méthodologique général se fonde sur :

1. un *input* précis et contraignant sous forme :
   (i)   d'un *modèle conceptuel* (d'une ontologie) du domaine d'expertise toujours amendable et
   (ii)  d'un modèle des *données textuelles* (lato sensu) ;
2. un *output* bien identifié et tout aussi contraignant sous forme d'un ensemble d'instructions anticipant le résultat souhaité d'une expertise ;
3. l'*examen* de volumes de *données textuelles qualifiées* conformément au modèle des données textuelles servant d'input (i.e. de données textuelles répondant aux *critères de qualité – auctoriale*, *informative*, *historique*, etc. ainsi que *structurale* à proprement parler – qui sont explicités dans le modèle des données textuelles).

Pour rester avec notre exemple du quechua comme objet de connaissance non seulement en linguistique, mais encore en littérature, en art, en archéologie et en anthropologie, les *projets de cartographie* répondent à une grande variété de questions, telles que :

–  Quelles sont les langues les plus proches/les plus éloignées du point de vue d'une caractéristique particulière ou d'un ensemble de caractéristiques ?
–  En considérant les caractéristiques décrivant le quechua, lesquelles sont les plus « populaires » (traitées le plus souvent) dans la littérature scientifique et lesquelles sont plutôt marginales et quelle est la particularité de cette échelle de popularité en comparaison avec la production scientifique sur d'autres langues ?
–  Quelles sont les convergences (et divergences) des points de vue qu'on rencontre dans la recherche travaillant sur une ou un ensemble de caractéristiques du quechua ?
–  Quels sont les scénarios d'évolution (en termes de politique linguistique, de transmission culturelle, …) plus ou moins probables de cette langue et/ou de la recherche sur cette langue ?
–  Peut-on localiser, dans la littérature scientifique, des problématiques qui ne sont pas/ne peuvent pas être identifiées dans le modèle conceptuel actuel du domaine d'expertise en référence auquel l'expertise sur cette langue est réalisée ?

Il s'agit ici, bien sûr, d'une simple liste ad hoc d'interrogations éventuelles auxquelles un travail de cartographie (d'analyse et de visualisation de données) peut apporter des réponses. Il est bien facile de s'imaginer d'autres cas de figure. Cela dit, des projets d'expertise de ce type, qui sont au cœur de toute recherche en sciences humaines et sociales, et qui constituent également un enjeu central pour la *gestion (scientifique et politique)* de la recherche, feront inévitablement un usage croissant des méthodes et des outils de l'intelligence artificielle (générative, prédictive et/ou agentive ; voir chapitre 1). La faisabilité et la qualité de ces projets seront ainsi profondément tributaires des avancées des recherches appliquées qui :

1. visent à « marier » l'intelligence artificielle *symbolique* (celle des bases de connaissance et des systèmes experts, travaillant avec des modèles conceptuels explicites) et l'intelligence artificielle *neuronale* (celle se basant principalement sur l'approche de l'apprentissage automatique) ;
2. investissent le domaine du design et du suivi des *prompts*, i.e. du design des instructions pour configurer correctement un agent conversationnel d'un modèle de langage (LLM) général (ou, éventuellement, d'un modèle déjà entraîné sur un corpus de données bien identifié) afin pour qu'il contribue d'une manière efficace et pertinente à la réalisation d'un projet d'expertise.

Le design de prompts relativement sophistiqués doit faire face, entre autres et comme déjà dit à deux problèmes récurrents que sont :

– d'une part, la *qualité des données* localisées et analysées et
– d'autre part, la *qualité de l'analyse elle-même* en fonction du modèle conceptuel d'un domaine de connaissance qui sert de standard aux *« pensées »* de l'agent conversationnel (i.e. au *parcours de sa pensée* ou « chain of thought », en anglais) qui sous-tend et guide son travail d'expertise.

D'où l'importance méthodologique dans le design d'un prompt de pouvoir recourir à un *modèle conceptuel* (à une ontologie) du *domaine de connaissance* qui ancre et contraint l'espace sémantique des actions et du raisonnement attendus d'un agent conversationnel correctement configuré. Le modèle conceptuel du domaine de connaissance précise les propriétés qui caractérisent les objets de connaissance qui en font partie.

Pour prendre l'exemple du quechua, le modèle servant d'input à un prompt peut identifier les valeurs de chaque propriété qui décrivent cette langue. Dans ce cas-là, il fixe, par exemple, le fait que le quechua fait partie de la macro-famille polyphylétique des langues amérindiennes, qu'il est une langue minoritaire, qu'il possède une structure phrastique SOV, etc. Mais, il peut également se contenter de fixer le *fait plus général* que pour rendre compte de la langue quechua, il faut identifier son appartenance à une famille linguistique, sa structure phrastique, son statut sociolinguistique, etc. Et, c'est à l'*agent conversationnel* (au « chatbot ») de trouver les informations appropriées à ce sujet, de les jauger, de les rédiger *tout en produisant une réflexion critique* expliquant sa démarche et ses choix.

Bien entendu, l'agent conversationnel peut également suggérer des caractéristiques non-retenues dans le modèle conceptuel servant d'input ce qui, éventuellement, peut conduire à une *modification* du modèle conceptuel du domaine (à son « ajustement », comme on dit, aux données analysées) et, donc, du graphe de connaissances dans l'écosystème de la base de connaissance.

A part du modèle conceptuel d'un domaine de connaissance (domaine tel que, pour reprendre notre exemple, celui des « Langues du monde »), il faut également disposer, répétons-le, d'un *deuxième* modèle pour contraindre et guider le travail d'expertise d'un agent conversationnel. Ce deuxième modèle spécifie explicitement les *caractéristiques structurales* des données textuelles lato sensu ainsi que les *critères de qualité* à prendre en compte pour sélectionner et analyser un corpus de données textuelles. En référence à ce modèle, l'agent conversationnel doit localiser et traiter, par exemple, uniquement des données textuelles d'un certain genre (scientifique), des données textuelles qui proviennent d'archives et d'entrepôts bien identifiés, des données textuelles qui comparativement à d'autres données textuelles antérieures, approfondissent (critiquent) les connaissances actuelles d'une caractéristique du quechua, etc.

L'*output attendu* du travail d'expertise d'un agent conversationnel se présente sous forme d'un modèle qui comprend, entre autres :

1. le *contenu attendu* de l'expertise ;
2. le *format* ou le *genre textuel* dans lequel le contenu doit prendre forme ;
3. les *modalités d'expression* et de *visualisation* selon lesquelles le contenu est rendu accessible.

Ainsi, on exige que le *contenu attendu* soit conforme au modèle conceptuel servant d'input. On fixe, entre autres, que le contenu produit par l'expertise doit *compléter, actualiser, problématiser, approfondir…* des connaissances déjà disponibles (sur une langue telle que le quechua, pour rester avec notre exemple). On peut aussi préciser le fait qu'il est attendu que l'expertise utilisera une *terminologie* précise et une *phraséologie* appropriée (à un contexte scientifique et académique), qu'elle produira, si pertinent, des *tableaux de synthèse*, des *diagrammes de visualisation* du contenu. Enfin, on peut également spécifier que l'output d'une expertise doit se présenter sous forme d'un simple *paragraphe*, d'un *chapitre*, d'un *dossier* à part entière, voire, si pertinent, sous forme de formats plus complexes tels que *cours*, *sites web*, etc.

Pour obtenir l'output désiré sous forme d'une expertise d'un objet de connaissance, il faut « programmer » l'agent conversationnel d'un modèle de langage général ou déjà pré-entraîné. C'est la partie de l'*ingénierie* de prompts qui se fait via des interfaces spécialisées qu'offrent les grands LLMs tels que ChatGPT, DeepSeek, Gemini, etc. Cette « programmation » se fait en partie en langage naturel – en français, par exemple. Mais, il s'agit ici tout de même d'un *langage contrôlé* qui doit se fonder sur une « vision » - une « théorie », si l'on veut - d'abord de l'output visé et des contraintes à prendre en compte pour y arriver.

Ajoutons encore qu'il convient de distinguer entre différents *types* ou *genres de prompts* pour bâtir un environnement permettant la conduite (semi-)automatique de projets d'expertise, par exemple, sur des objets de connaissance précis en études aréales. Cette distinction entre différents types ou genres de prompts se fait plutôt au niveau des instructions qui les composent. Tout en sachant que les recherches avancent ici d'une manière extrêmement rapide, mentionnons à titre d'exemple :

- les *prompts génériques* : des prompts qui intègrent des suites d'instructions (hiérarchisées) qu'on n'utilise pas une seule fois, mais à maintes reprises (sans ou avec des modifications/adaptations locales de certaines instructions), par exemple, pour expertiser les objets d'un domaine de connaissance tel que celui intitulé « Langues du monde » qui fait partie de la base de connaissance LaCAS ;
- les *prompts généraux* : des prompts qui intègrent des instructions ou des suites d'instructions qu'on peut retrouver dans différents prompts, voire, à la limite, dans tous les prompts utilisés, par exemple, en études aréales pour alimenter une base de connaissance ;
- les *prompts contextuellement ancrés* : des prompts qui intègrent des instructions qui les contraignant à travailler exclusivement dans un espace sémantique explicité par un ou plusieurs modèles qui leur servent d'input ;
- les *prompts simulant un raisonnement* : des prompts qui intègrent des instructions les contraignant à fournir des explications de leur choix et de leur « travail » qu'ils sont en train de réaliser ;
- les prompts d'ajustement : des prompts qui intègrent des instructions pour ajuster, adapter un modèle de référence (servant d'input au travail d'un agent conversationnel) aux données analysées ;
- les *prompts réalisant une série de tâches :* des prompts qui intègrent des instructions les contraignant à suivre un plan d'actions, éventuellement de hiérarchiser des actions et des tâches selon une maxime d'importance, etc. ;
- les *prompts répétant une* ou *une série actions* : des prompts qui intègrent des instructions les contraignant de répéter certaines actions pendant une période déterminée ou indéterminée ;
- les *prompts exhibant un certain comportement face à un usager (humain ou artificiel)* : des prompts qui intègrent des instructions les contraignant à respecter certaines règles d'interaction avec des usagers humains et/ou d'autres agents artificiels.

En contemplant ces divers types de prompts et sans pouvoir approfondir ici leur structure et leur fonctionnement, on peut cependant s'imaginer d'éventuelles conséquences d'un mariage – évoqué au début de ce chapitre – entre les techniques de l'intelligence artificielle symbolique et celles de l'intelligence artificielle dite neuronale[19].

L'introduction des techniques génératives, prédictives et agentives issues de l'intelligence artificielle neuronale dans l'écosystème d'une base de connaissance telle que celui de LaCAS, présenté ci-dessus (chapitre 5), pourra « animer » la base de connaissance, la transformer en une sorte de – métaphoriquement parlant – organisme artificiel « capable » de s'enrichir d'une manière autonome, de s'adapter aux données issues de son milieu, de se différencier en un *système reparti* de bases de connaissance plus spécialisées, de prendre de décision pour lancer de projets d'expertise sur des objets de connaissance particulier, de renseigner d'une manière sélective sur ses domaines et objets de connaissance, d'interagir avec des écosystèmes d'autres bases de connaissance.

Une base de connaissance d'un tel genre tout à fait nouveau s'appuiera sur le travail d'une *communauté d'agents artificiels* spécialisés avec laquelle elle constitue un écosystème, comprenant également un environnement interne, des pratiques et des activités, des domaines de spécialité et se distinguant par ses dynamiques téléologiques propres. Chaque agent contribuera à la « vie » de la base de connaissance et à son évolution » selon le *prompt* – selon les séries d'instructions – qui le définit. C'est une perspective qui est, bien entendu, à l'heure actuelle, encore très spéculative, mais certainement pas dépourvu d'arguments tout à fait sérieux.

Néanmoins, un tel écosystème artificiel reste intrinsèquement « humain », dans la mesure où il est le produit de l'imagination humaine, et donc limité par notre langage. Mais on peut, en effet, se demander dans quelle mesure il resterait un simple outil au service et sous le contrôle des humains s'il était un jour réalisé.

---

[19] Cf. au sujet de cette problématique Wang, W., Yang, Y., & Wu, F. (2022). Towards data- and knowledge-driven artificial intelligence: A survey on neuro-symbolic computing. *arXiv preprint arXiv:2210.15889*.

## PETIT GLOSSAIRE

- *Base de connaissance :* Socle structuré d'objets et de propriétés (faits) activés par un raisonneur pour produire réponses, décisions ou analyses ; cœur d'un écosystème de traitement et de publication, conçu ici comme un *milieu sémiotique.*

- *Graphe de connaissances :* Représentation en nœuds/arêtes des entités et relations d'un domaine, permettant requêtes, inférences et mises à jour raisonnées.

- *Web sémantique (RDF/OWL) :* Modélisation en graphes (triplets) et langages ontologiques pour décrire, relier et inférer sur des entités d'un domaine (classes, propriétés, contraintes).

- *Données ouvertes liées (LOD) :* Référentiels interconnectés (p. ex. Wikidata) servant d'alignements sémantiques, de points d'ancrage et de hubs d'informations réutilisables.

- *I.A. symbolique :* Approche fondée sur des modèles explicites (ontologies, règles logiques) et des moteurs d'inférence ; matrice historique des systèmes experts et des bases de connaissance.

- *Systèmes experts :* Systèmes symboliques associant base de règles et base de faits pour résoudre des problèmes délimités avec traçabilité des justifications.

- *I.A. neuronale :* Approche connexionniste utilisant des représentations vectorielles et l'entraînement de réseaux (génératif, prédictif, agentif), sans ontologie explicite du domaine.

- *Modèles d'I.A. générative :* Modèles capables de produire texte, code, image, etc. ; dans l'article, ils sont contraints par l'ontologie et des prompts contextualisés pour animer la base sans en rompre les règles.

- *Approche model-driven :* Méthodes guidées par le modèle conceptuel (ontologie, schémas) : alignement, validation, suggestion de propriétés, contrôle des sorties.

- *Approche data-driven :* Méthodes guidées par les données (signaux, embeddings, statistiques) : découverte de motifs, complétude, priorisation et alertes.

- *Ingénierie de prompts :* Conception d'instructions hiérarchisées et contextualisées (alignées sur l'ontologie et le modèle des données) pour piloter des agents génératifs.

- *Écosystème d'agents :* Ensemble coordonné d'agents spécialisés (moissonnage, normalisation, alignement, rédaction, contrôle) qui « animent » la base en respectant ses contraintes symboliques.

- *Agentivité non-humaine :* Capacité d'agir distribuée dans des artefacts/organismes non humains, reconfigurant les cadres de l'interprétation et de l'interaction.

- *Études aréales :* Approches pluridisciplinaires centrées sur des aires géopolitiques, linguistiques, culturelles ou historiques, instrumentées ici par une base de connaissance.

- *LaCAS / Okapi :* Plateforme sémantique de l'Inalco (LaCAS) pour collecter/traiter/republier des données d'études aréales, s'appuyant sur le logiciel Okapi (Ina).